\documentclass[a4paper,fleqn,usenatbib]{mnras}

\usepackage{graphicx}
\usepackage{pdfpages}
\usepackage{subcaption}
\usepackage{float}
\usepackage{mwe}
\usepackage{xcolor,colortbl}

\usepackage[T1]{fontenc}
\usepackage{ae,aecompl}

\usepackage{graphicx}	
\usepackage{amsmath}	
\usepackage{amssymb}	
\usepackage{threeparttable}
\usepackage{adjustbox}

\title[Star formation in MB-C]{Resolved star formation in the metal poor star-forming region Magellanic Bridge C}

\author[V. M. Kalari et al.]{
Venu M. Kalari$^{1,2}$\thanks{E-mail: vkalari@gemini.edu},
Monica Rubio$^{2}$,  
Hugo P. Salda{\~n}o$^{2, 3}$, and
Alberto D. Bolatto$^{4}$
\\
$^{1}$Gemini Observatory, Southern Operations Center, c/o AURA, Casilla 603, La Serena, Chile\\
$^{2}$Departamento de Astronomia, Universidad de Chile, Casilla 36-D, Santiago, Chile\\
$^{3}$Observatorio Astron\'omico, Universidad Nacional de C\'ordoba, C\'ordoba, Argentina\\
$^{4}$Department of Astronomy, University of Maryland, College Park, MD\,20742, USA\\
}

\date{Accepted  . Received  ; in original form  }

\pubyear{2015}

\begin{document}
\label{firstpage}
\pagerange{\pageref{firstpage}--\pageref{lastpage}}
\maketitle

\begin{abstract}
Magellanic Bridge C (MB-C) is a metal-poor ($\sim$1/5\,$Z_{\odot}$) low-density star-forming region located 59\,kpc away in the Magellanic Bridge, offering a resolved view of the star formation process in conditions different to the Galaxy. From Atacama Large Millimetre Array CO\,(1-0) observations, we detect molecular clumps associated to candidate young stellar objects (YSOs), pre-main sequence (PMS) stars, and filamentary structure identified in far-infrared imaging. YSOs and PMS stars form in molecular gas having densities between 17--200\,$M_{\odot}$\,pc$^{-2}$, and have ages between $\lesssim$0.1--3\,Myr. YSO candidates in MB\,-C have lower extinction than their Galactic counterparts. Otherwise, our results suggest that the properties and morphologies of molecular clumps, YSOs, and PMS stars in MB\,-C present no patent differences with respect to their Galactic counterparts, tentatively alluding that the bottleneck to forming stars in regions similar to MB-C is the conversion of atomic gas to molecular. 
\end{abstract}




\section{Introduction}

In the current paradigm of star formation, turbulence and self-gravity play governing roles in the processes determining the eventual conversion of molecular gas to stars across vast physical scales (McKee \& Ostriker 2007). These star-forming processes can be qualitatively studied based on scaling relationships between quantities, and the properties and morphologies of molecular clouds and the newly forming stars within them (e.g. Hartmann, Herczeg \& Calvet 2016; Evans, Heiderman \& Vutisalchavakul 2014; Elmegreen 2011; Bergin \& Tafalla 2007; Lada \& Lada 2003). Whether these principal star formation processes differ across the variety of interstellar and Galactic environments observed in nature, over the physical scales of galaxies and giant molecular clouds, to individual stars is yet to be quantified exhaustively (Kennicutt \& Evans 2012). As a unified theory of star formation must specify the dominant processes across such diverse environments, it becomes necessary to obtain observational constraints on star formation processes across a variety of environments, to inform theory. 

Significant strides have been made at the observational front, intensifying with the improved capabilities of the current generation of facilities. Observations of molecular clouds (Evans et al. 2009) in the Milky Way have shown that nearly all star formation is localised into molecular clumps (with loosely defined density thresholds e.g. Heidermann et al. 2010; Lada, Lombardi \& Alves 2010; although these are contested, e.g. Burkert \& Hartmann 2013), which form filamentary structures (Andr{\'e} 2017) before being nearly universally associated with young, forming stars. It remains open whether the relationship between molecular cloud clumps and young protostars is similar in differing environments. Stars are built from molecular gas with low efficiencies, with the density of nascent clouds crucial in determining the densities of the formed stellar associations and resultant star forming efficiency (Longmore et al. 2014; Evans et al. 2014; Stahler \& Palla 2005). Whether the properties of stars thus formed (e.g. the stellar initial mass function Bastian, Covey \& Meyer 2010; forming efficiencies, e.g. Kennicutt \& Evans 2012; Genzel et al. 2010; Evans et al. 2009) are similar irrespective of the properties of the nascent molecular cloud, and the local environment (stellar and gas density, ionising radiation field, metallicity) is debatable.

\begin{figure*}
\center 
\includegraphics[width=16cm]{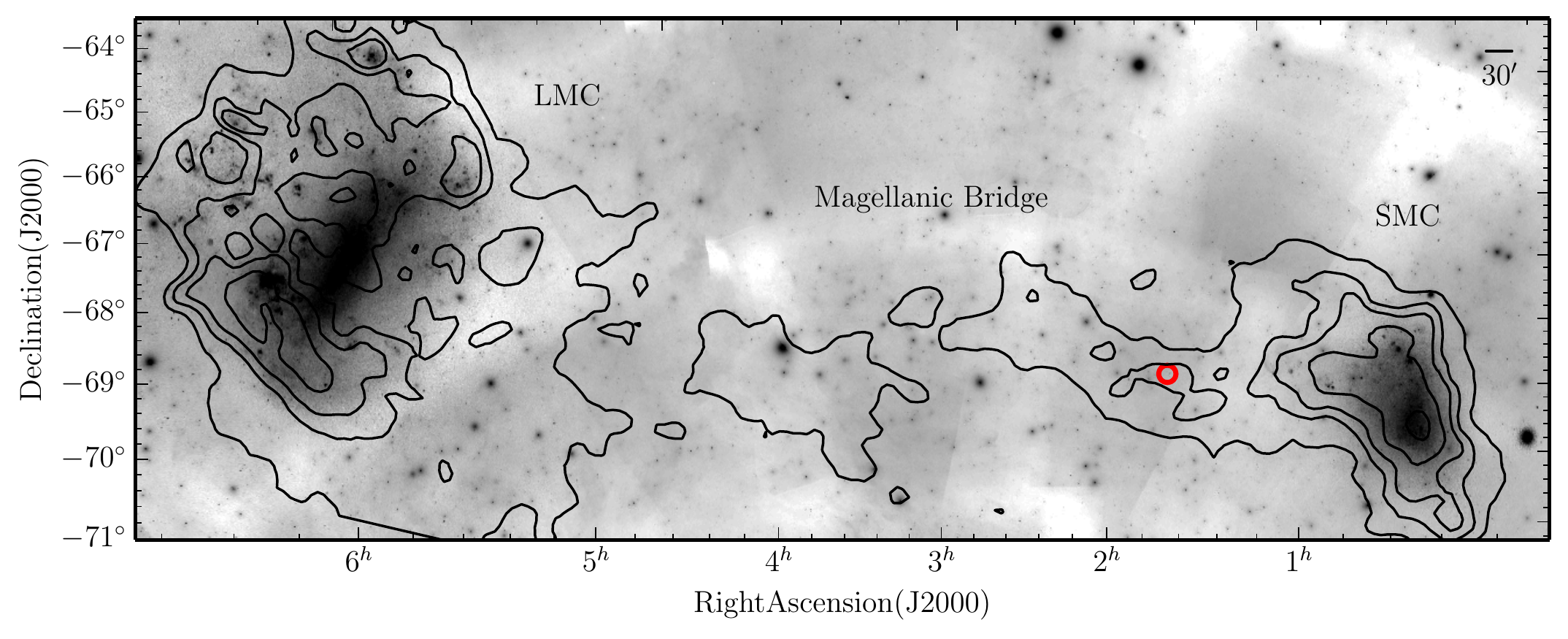}
\center
\caption{The position of MB-C in the sky from an archive Digital Sky Survey image, along with the position of the Large and Small Magellanic Clouds labelled and seen as overdensities in the optical image. Overlaid are \ion{H}{I} contours from the Parkes All Sky Survey (Meyer et al. 2004) at levels of 1, 10, 25, 50 and 100 counts. MB-C is marked by the solid circle, showing its relative isolation from the centres of the SMC ($\sim$3kpc) and the LMC ($\sim$12\,kpc). North is up and east is to the left.}
\label{fig:allfigs}
\end{figure*}

In this paper we study the low-density low metallicity ($Z$) star-forming region Magellanic Bridge\,-C (hereafter MB\,-C) located in the intergalactic region Magellanic Bridge (Irwin, Kunkel \& Demers 1985). The Magellanic Bridge (from R.A. between $\sim$1$^h$30$^m$ to 5$^h$) is a filament of \ion{H}{I} lying at a heliocentric distance of $\sim$50-60\,kpc evolving gravitationally under the influence of the two dwarf Galaxies located at its extremities, the 1/2\,$Z_{\odot}$ Large Magellanic Cloud (LMC), and the 1/5\,$Z_{\odot}$ Small Magellanic Cloud (SMC). MB\,-C (at an R.A.$\sim$2) lies slightly east of the SMC wing (1$^h\lesssim$\,R.A.\,$\lesssim$1$^h$30$^m$), and is pictured in Fig.\,1. The region is centred on the molecular cloud C (see Fig.\,1) first identified in CO by Mizuno et al. (2006). MB\,-C lies encompassed within the high \ion{H}{I} density region of the Bridge, where $N_{{\rm HI}}\sim10^{21}\,$cm$^{-2}$--3$\times10^{21}\,$cm$^{-2}$ (between R.A.$\sim$1$^h$30$^m$ to 2$^h$30$^m$; see also Fig. 1 in Mizuno et al. 2006), where CO molecular clouds had been first detected in the Bridge. An intermediate mass $\sim$20\,Myr cluster ($\sim10^3\,M_{\odot}$) NGC\,796 (Kalari et al. 2018a) lies approximately 20\,pc north of MB\,-C, containing many early-B type stars suggesting recent star formation within the region. The distance to that cluster is adopted as the distance to MB\,-C in this paper. The metallicity estimated from stellar abundances of B-type stars in the region near MB\,-C is around 1/5\,$Z_{\odot}$ (Lee et al. 2005), while both gas-phase and stellar abundances in the enveloping inter-cloud region (including the high \ion{H}{I} density region) suggest a lower metallicity of 1/10\,$Z_{\odot}$ (Rolleston et al. 1999; Dufton et al. 2008; Lehner et al. 2008). As the metallicity within the region has hitherto been unmeasured, we assume for the remainder of our work a metallicity of 1/5\,$Z_{\odot}$ to be representative for the region. The metallicity of the region represents conditions similar to star forming disc galaxies at redshifts\,$\sim$\,4 (Kewley \& Kobulnicky 2005).

The molecular cloud itself contains three previously identified stellar associations, BS\,216, BS\,217, and WG\,8 (see Bica et al. 2008), although their stellar properties are not determined in the literature. Distinct H$\alpha$ blobs have been identified in MB\,-C (Muller \& Parker 2007), but their relation to massive stars formed there (as opposed to having travelled there from the SMC) has not yet been verified. However, early B-type stars in the region have been identified by Chen et al. (2014). Chen et al. (2014) performed a detailed mid-infrared (mIR) star formation study of the Magellanic Bridge using {\it Spitzer} space telescope imaging covering 3.6--160$\mu$m. They found that MB-C has no massive young stellar objects (mYSOs), although six faint young stellar objects (YSOs) are detected. These YSOs were comparatively less embedded than their LMC counterparts. They concluded that the high mass of molecular cloud C ($\sim$7$\times$10$^3$\,$M_{\odot}$; from Mizuno et al. 2006) has resulted in a more clustered mode of star formation than in other regions of the Magellanic Bridge. 

In summary, from the literature, MB\,-C contains known CO molecular clouds, and YSOs in various stages of evolution. It is a low-density, inter-galactic, metal-poor star-forming region. Its proximity (at 59\,kpc) offers a unique opportunity to study resolved star formation from the molecular cloud to the pre-main sequence phase, in physical conditions resembling that of more distant dwarf galaxies which cannot be resolved to the level afforded here, thereby allowing a unique window into how star formation proceeds in a metal-poor regime. 

We present Atacama Large Millimeter Array (ALMA) CO\,(1-0) observations towards MB-C to trace the molecular clumps forming stars. In addition, we utilise mIR imaging from the {\it Spitzer} space telescope to identify and characterise the extremely young YSOs. The more evolved optically visible pre-main sequence stars are identified using deep optical imaging (including H$\alpha$). In our study, we focus on the properties and morphologies of the molecular clouds and young forming stars in MB-C. 


This paper is organised as follows. Section 2 presents ALMA imaging of the molecular clouds. Section 3 lists the results of identifying YSOs using mIR imaging. In Section 4, adaptive optic observations in optical broadband and H$\alpha$ narrowband are presented to identify accreting YSOs. In Section 5, a discussion of the results are presented. Finally, in Section 6 our conclusions are presented.

\begin{figure}
\center 
\includegraphics[width=8cm]{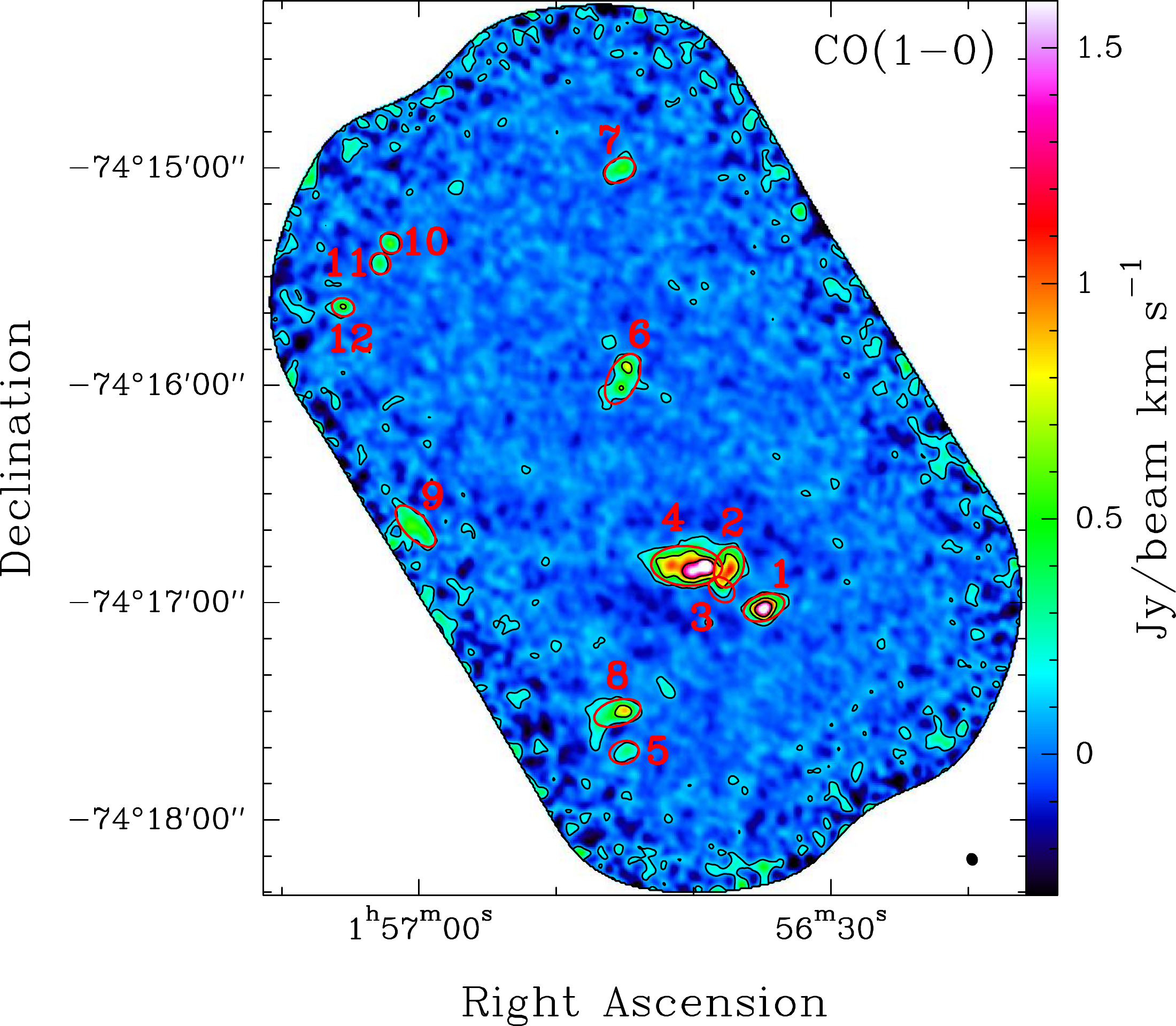}
\center
\caption{CO\,(1-0) map of MB-C integrated over the 175.5--185.0\,km\,s$^{-1}$ velocity range. The contours have values of 0.13, 0.63 and 1.13 Jy\,beam$^{-1}$\,km\,s$^{-1}$, with the r.m.s.\,=\,0.027 Jy\,beam$^{-1}$. In the bottom right corner, the beam size is shown. The major axis of the beam is 3.6$\arcsec$. Ellipses indicate the size and location of each identified clump numbered according to Table\,1.}
\label{comap}
\end{figure}

\section{ALMA CO\,(1-0) observations of molecular clumps }
Star formation takes place almost exclusively within molecular clouds, which are comprised primarily of molecular hydrogen (H$_2$). However, the lack of a permanent dipole moment in H$_2$ means that often, carbon monoxide (CO), the second most abundant component in molecular clouds is used to study them. CO exhibits multiple transitions detectable primarily at sub-mm wavelengths, allowing one to trace both the amount and density of the cold molecular gas from which stars form. 

\subsection{Observations}

CO\,(1-0) observations of the Magellanic Bridge were carried out using the ALMA telescope on 2016 January 22 under the ALMA Program 2015.1.1013.S (PI: M. Rubio). The source was mapped with ALMA 12-m antennas using the Band 3 receiver (84--116 GHz) at a beam size of 3.6$\arcsec \times$3.3$\arcsec$ ($\sim$0.9\,pc at the distance to MB-C), and a spectral resolution of 0.5\,km\,s$^{-1}$. The map covers a field of view (FoV) of 4.1$\arcmin \times$2.6$\arcmin$ ($\sim$75$\times$47\,pc at 59\,kpc) with a PA\,=\,$+$53$^{\circ}$. The average value of the r.m.s. noise ($\sigma_{\rm rms}$) is 0.027\,Jy\,beam$^{-1}$ (0.21\,K), and the maximum recoverable scale (MRS) is 25.4$\arcsec$.
The data was reduced using the Common Astronomy Software Application (CASA; McMullin et al. 2007). 

\subsection{Clump identification and analysis}

In Fig.\,2, the distribution of CO clouds towards MB-C (integrated over the velocity range 175.5--185\,km\,s$^{-1}$) is shown. Compact emission is detected, and the {\textsf {CPROPS}} algorithm of Rosolowsky \& Leroy (2006) is utilised to compute their properties. Cloud identification is based on the construction of a mask that only considers emission above 4$\sigma_{\rm rms}$ and edge threshold of about 2.0--1.5$\sigma$. Once the clouds are identified, we use the modified {\textsf {CLUMPFIND}} algorithm (setting the keyword {\textsf {/ECLUMP}}) to apply the decomposition process on the CO distribution. The central velocity  ($V_{\rm lsr}$), radius ($R$), velocity dispersion ($\sigma$), the CO luminosity ($L_{\rm CO}$), and the virial mass ($M_{\rm vir}$) for each clump was estimated and listed in Table 1. These parameters are estimated through the moment method, using the distribution of emission in a cloud within a position-position-velocity data cube without assuming a previous functional form for the cloud. A Gaussian profile is assumed to fit the CO line. The estimated parameters are corrected for sensitivity bias by extrapolating the properties to those we would expect to measure with perfect sensitivity (i.e. $\sigma_{\rm rms}$\,=\,0\,K). The clump radius and $V_{\rm lsr}$ are deconvolved to be corrected for finite spatial and spectral resolutions. All clouds are well resolved by {\textsf {CPROPS}} with the exception of the weakest ones, clouds 10, 11, and 12. These clouds are included in Table 1, but their sizes correspond to non-deconvolved sizes. 

\begin{figure}
\center 
\includegraphics[width=8cm]{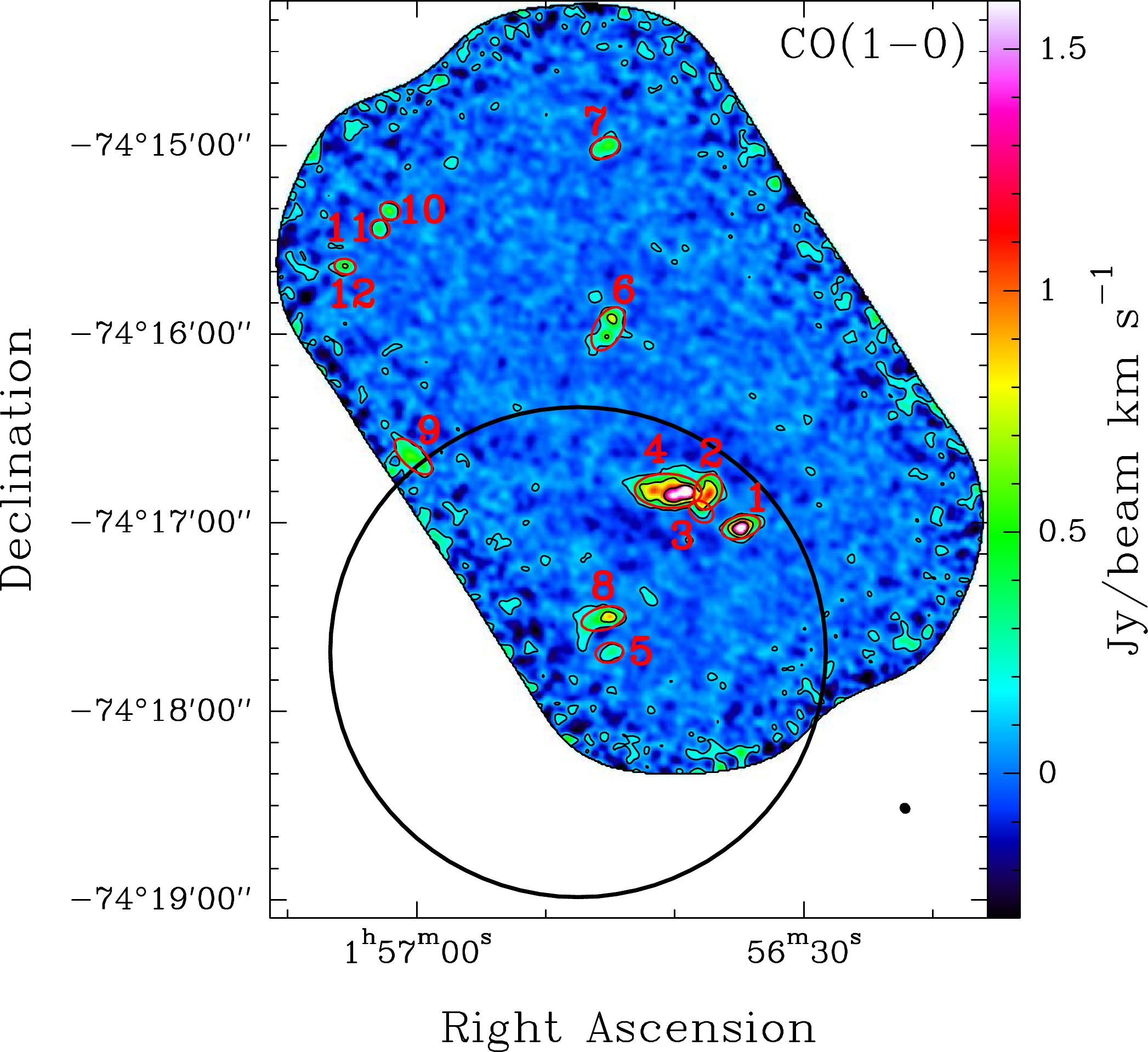}
\includegraphics[width=8cm]{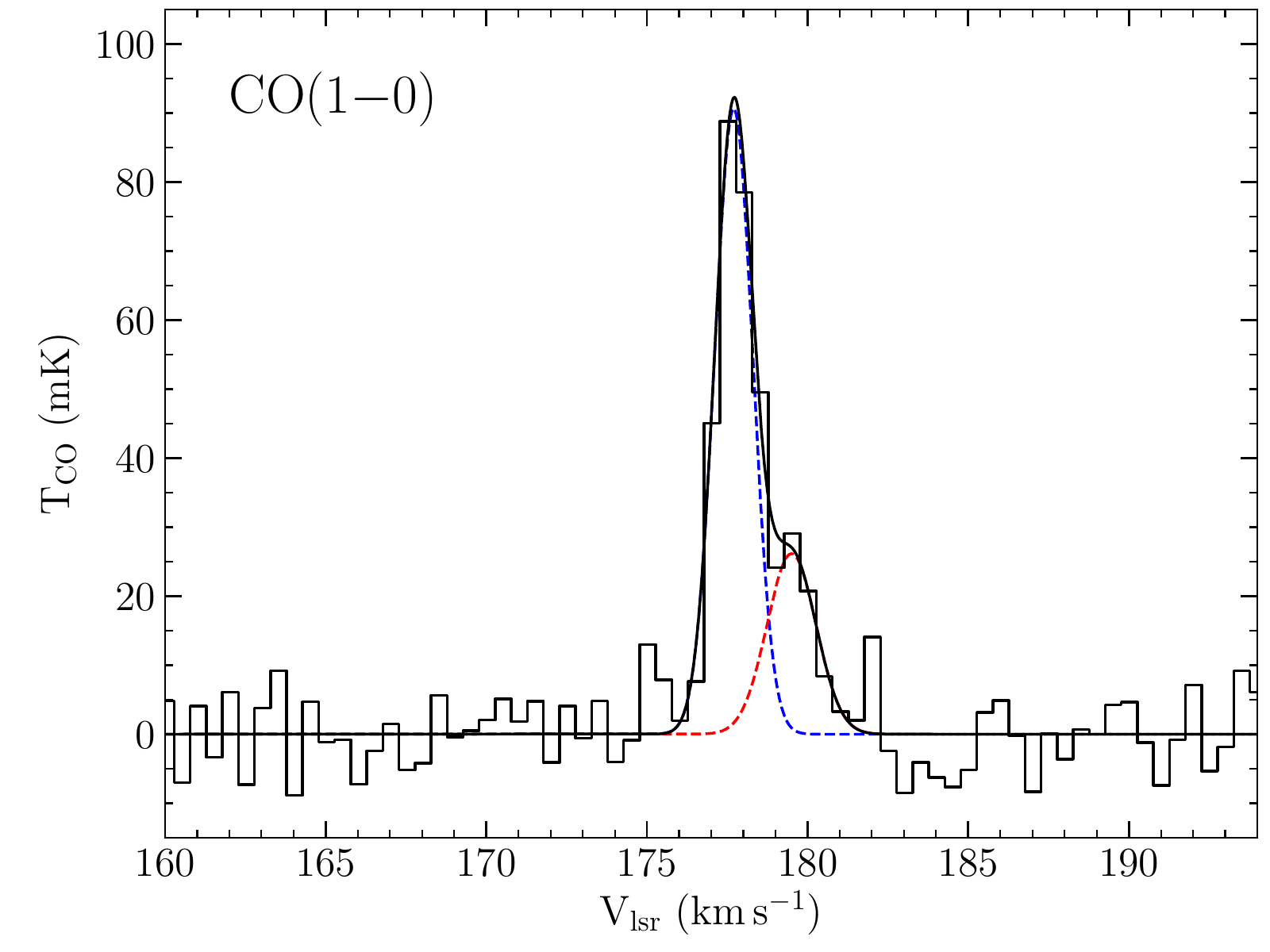}
\center
\caption{{\it Top panel:} ALMA CO\,(1-0)  map  of  MB-C from Fig. 2, with detected molecular clouds from ALMA are labelled. The beam of the single-dish NANTEN observations from Mizuno et al. (2006) is given by the solid line. {\it Bottom panel:} CO\,(1-0) integrated spectra from ALMA data taken from the region which partially coincides with the NANTEN beam (2.6$\arcmin$). The Gaussian fit (solid black) with two dashed components (solid blue and red) are also shown.}
\label{comparealma}
\end{figure}

ALMA observations are not corrected for total power. To estimate if we miss any significant contribution diffuse CO emission, we compared the CO luminosities of the clouds in the ALMA map with the luminosity of the region reported by Mizuno et al. (2006) using NANTEN single-dish observations. For this comparison, we need to consider that the region mapped with ALMA coincides only partially with the NANTEN telescope observations, which consist of one pointing with a 2.6$'$ beam offset relative to the ALMA map. 
There are seven ALMA CO clouds in our map that contribute to the emission in the NANTEN beam, clouds 1--5, 8, and 9 as can be seen in Fig.\,\ref{comparealma}, which shows the NANTEN beam overlaid on the ALMA map. We added the total luminosity of these clouds corrected for the NANTEN beam pattern to obtain a total luminosity $L_{\text{CO}}=\,$181.1$\pm$27\,K\,km\,s$^{-1}$\,pc$^{2}$. We determine the NANTEN CO luminosity using the reported $I_{\text{CO}}$ of 140\,mK\,km\,s$^{-1}$ by Mizuno et al. (2006), and obtain a NANTEN CO luminosity\footnote{To calculate the NANTEN CO luminosity, the Gaussian beam size was determined following $\pi/\ln(2)\,\times({\textrm{FWHM}}/2)^2$, and is 2256 \,pc$^2$.} of 315.5\,K\,km\,s$^{-1}$\,pc$^{2}$. The area of the ALMA MB-C region overlapping with the NANTEN FWHM circle is 902.2$\pm$8\,pc$^2$, while the total area of the FWHM circle is 1564\,pc$^2$. Thus, 58 per\,cent of  NANTEN observed luminosity is recovered by ALMA with a map that covers 58 per\,cent of the FWHM of the NANTEN beam. This is quite remarkable, and suggests it is unlikely that more than $\sim$20\,per\,cent of the flux is in a diffuse component resolved out by the interferometer.

We also fit the ALMA spectrum integrated over the FWHM of the NANTEN beam with two Gaussian components.  The most intense component is centred at $V_{\text{lsr}}$=\,177.7$\pm$0.1\,km\,s$^{-1}$, and the second weaker component is observed at $V_{\text{lsr}}$=\,179.5$\pm$0.5\,km\,s$^{-1}$. The linewidth of the ALMA CO integrated spectrum is 2.4\,km\,s$^{-1}$, while each velocity component shows a linewidth of 1.4$\pm$0.2\,km\,s$^{-1}$, and 1.8$\pm$0.8\,km\,s$^{-1}$, respectively. Table 1 shows the individual linewidth for each of the CO clouds,
expressed as velocity dispersions, and all of them are narrow and of the same order, less than 2\,km\,s$^{-1}$. It is noteworthy that the CO linewidth (1.4$\pm$0.2\,km\,s$^{-1}$) of the main component in the integrated ALMA spectrum is similar to the CO linewidth of 1.5\,km\,s$^{-1}$ reported by Mizuno et al. (2006), and the second, fainter velocity component of the integrated profile is at the level of the noise of the NANTEN spectrum.

\subsection{Clump parameters}

Twelve clumps in MB-C having signal-to-noise greater than 13 at the peak are detected. Their locations, and estimated properties are given in Table\,1. The virial mass ($M_{\rm vir}$) is calculated assuming a spherical cloud with a 1/$r$ density profile (MacLaren, Richardson \& Wolfendale 1988). We calculate the clump surface density ($\Sigma_{\rm H_2}$) based on their virial masses. The corresponding surface density r.m.s. is calculated following Rosolowsky \& Leroy (2006), and assuming a distance of 59\,kpc is 0.46\,$M_{\odot}$\,pc$^{-2}$ (this corresponds to the r.m.s. beam sensitivity of 0.027\,$Jy$\,beam$^{-1}$). The location, and size of all clumps are indicated in Fig.\,\ref{comap} by red ellipses. The radii and masses of the observed entities indicate that they are molecular clumps in the commonly used definition of the term (Bergin \& Tafalla 2007). 

Most molecular clouds, and cloud clumps having radii between 2-15\,pc (Bergin \& Tafalla 2007), are  gravitationally bound being close to a virialized state. They follow the relations first observed by Larson (1981), highlighting a given cloud's gravoturbulent state, thereby relating them to star formation. Most smaller clump structures are often observed not to obey Larson's relation (McKee \& Ostriker 2007), but observational difficulties may partly be responsible in some cases.

To examine the relation of these molecular clumps to star formation, we examine their properties against those predicted by Larson's laws for clouds in  self-gravitational equilibrium. We also compare our clumps to molecular cloud cores of similar sizes in our Galaxy identified in the seminal study by Heyer et al. (2009), and to molecular clumps found in differing star-forming environments, namely the low and high gas surface density regimes (see Kennicutt \& Evans 2012 for a discussion on the boundaries between the regimes). In the high-density star-forming regime, we compare our clumps properties to those observed in high-mass star-forming region of 30 Doradus (north-east to the superstar cluster R\,136, in the cloud 10 region of the nebula) of the LMC (Nayak et al. 2016). In the low density star-forming regime, we compare our clumps to cloud cores found in the outer Galaxy (Heyer, Carpenter \& Snell 2001). In these data sets, we only compare clumps within the radii range (0.2--2\,pc) of our clumps.

\begin{figure}
\center 
\includegraphics[width=8cm]{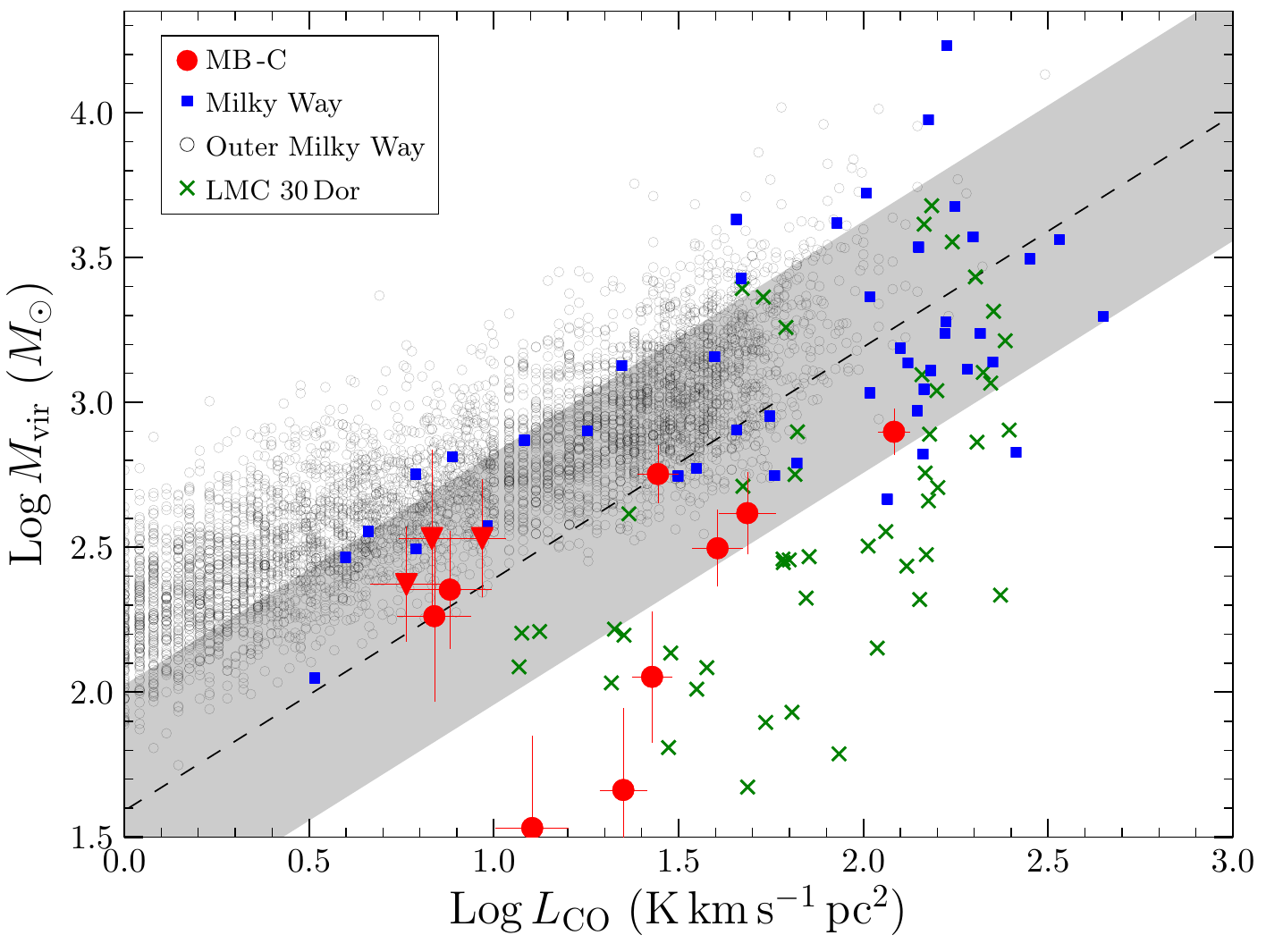}
\center
\caption{The variation of CO luminosity $L_{\rm CO}$, and the the virial mass of MB\,-C molecular clumps are shown as filled circles. Unresolved clumps are indicated by inverted carets. The approximate relation for Galactic molecular clumps of $M_{\rm vir}$=39$L_{\rm CO}^{0.81}$, according to Solomon et al. (1989) is also shown as a dashed line. The shaded portion lines represents a departure of 0.5\,dex for the virial mass from this relationship. The properties of clumps in the Milky Way (Heyer et al. 2009) are shown as squares. The crosses represent the values for clumps in the prototype starburst 30 Doradus in the LMC from Nayak et al. (2016), and the circles represent the properties of clumps in the outer Milky Way from Heyer et al. (2001).  }
\label{larson}
\end{figure}

\begin{table*}
\begin{adjustbox}{width=0.9\textwidth}
\begin{threeparttable}
\label{cotable}
	\centering
	\caption{Properties of CO\,(1-0) molecular clumps detected by ALMA observations$^a$}

\begin{tabular}{lccccccccc}
\hline
  \multicolumn{1}{c}{ID} &
  \multicolumn{1}{c}{Right} &
  \multicolumn{1}{c}{Declination} &
  \multicolumn{1}{c}{$V_{\rm lsr}$} &
  \multicolumn{1}{c}{$R$} &
  \multicolumn{1}{c}{$\sigma$} &
  \multicolumn{1}{c}{$L_{\rm CO}$} &
  \multicolumn{1}{c}{$M_{\rm vir}$} &
  \multicolumn{1}{c}{$\Sigma_{\rm H_2}$$^{b}$} \\
    \multicolumn{1}{c}{} &
  \multicolumn{1}{c}{Ascension} &
  \multicolumn{1}{c}{} &
  \multicolumn{1}{c}{(km\,s$^{-1}$)} &
  \multicolumn{1}{c}{(pc)} &
  \multicolumn{1}{c}{(km\,s$^{-1}$)} &
  \multicolumn{1}{c}{(K\,km\,s$^{-1}$\,pc$^{2}$)} &
  \multicolumn{1}{c}{($M_{\odot}$)} &
  \multicolumn{1}{c}{($M_{\odot}$\,pc$^{-2}$)}\\
\hline\hline
  1 & 01$^h56^m$34$^s$86 & $-$74\degr17\arcmin01\arcsec5 & 179.6$\pm$0.5 & 1.1$\pm$0.2 & 0.6$\pm$0.1 & 48.5$\pm$8.7 & 415$\pm$137 & 109$\pm$54\\
  2 & 01$^h56^m$37$^s$42 & $-$74\degr16\arcmin50\arcsec3 & 177.5$\pm$0.5 & 1.2$\pm$0.2 & 0.5$\pm$0.1 & 40.3$\pm$6.4 & 314$\pm$96 & 70$\pm$31\\
  3 & 01$^h56^m$37$^s$95 & $-$74\degr16\arcmin56\arcsec6 & 178.8$\pm$0.6 & 0.6$\pm$0.2 & 0.6$\pm$0.1 & 7.6$\pm$2.0 & 226$\pm$106 & 200$\pm$160\\
  4 & 01$^h56^m$40$^s$51 & $-$74\degr16\arcmin50\arcsec0 & 177.9$\pm$0.6 & 2.1$\pm$0.1 & 0.6$\pm$0.1 & 121.0$\pm$12.1 & 792$\pm$147 & 57$\pm$12\\
  5 & 01$^h56^m$45$^s$06 & $-$74\degr17\arcmin41\arcsec5 & 177.2$\pm$0.5 & 0.7$\pm$0.2 & 0.5$\pm$0.2 & 6.9$\pm$1.6 & 183$\pm$125 & 120$\pm$106\\
  6 & 01$^h56^m$45$^s$17 & $-$74\degr15\arcmin58\arcsec3 & 177.7$\pm$0.5 & 1.5$\pm$0.2 & 0.6$\pm$0.1 & 27.8$\pm$3.6 & 566$\pm$131 & 80$\pm$28\\
  7 & 01$^h56^m$45$^s$39 & $-$74\degr15\arcmin00\arcsec7 & 177.8$\pm$0.5 & 0.8$\pm$0.3 & 0.2$\pm$0.1 & 12.7$\pm$2.9 & 34$\pm$25 & 17$\pm$17\\
  8 & 01$^h56^m$45$^s$56 & $-$74\degr17\arcmin30\arcsec6 & 177.5$\pm$0.5 & 1.2$\pm$0.2 & 0.3$\pm$0.1 & 26.8$\pm$3.4 & 113$\pm$59 & 25$\pm$15\\
  9 & 01$^h57^m$00$^s$29 & $-$74\degr16\arcmin39\arcsec0 & 177.7$\pm$0.5 & 1.1$\pm$0.2 & 0.2$\pm$0.1 & 22.4$\pm$3.3 & 46$\pm$30 & 12$\pm$9\\
  10$^b$ & 01$^h57^m$02$^s$07 & $-$74\degr15\arcmin20\arcsec8 & 176.9$\pm$0.6 & 0.9$\pm$0.2 & 0.6$\pm$0.2 & 6.8$\pm$1.4 & 340$\pm$240 & 133$\pm$110\\
  11$^b$ & 01$^h57^m$02$^s$86 & $-$74\degr15\arcmin26\arcsec5 & 178.2$\pm$0.5 & 0.9$\pm$0.2 & 0.5$\pm$0.1 & 5.8$\pm$1.3 & 236$\pm$108 & 93$\pm$60\\
  12$^b$ & 01$^h57^m$05$^s$57 & $-$74\degr15\arcmin38\arcsec5 & 177.4$\pm$0.6 & 0.9$\pm$0.3 & 0.6$\pm$0.1 & 9.3$\pm$1.4 & 340$\pm$160 & 133$\pm$109\\
\hline
\end{tabular}
\begin{tablenotes}{
\item[] $^{a}$ All uncertainties quoted are estimated using {\textsf {CPROPS}}. $^{b}$ The radius of these sources are not deconvolved, and not used to estimate the $X_{\rm CO}$ conversion factor. $^c$The surface density is calculated from H$_2$ virial mass.}
\end{tablenotes}
\end{threeparttable}
\end{adjustbox}
\end{table*}

We plot the virial mass against the gas luminosity in CO, and display it along with the empirical relation for Galactic molecular clouds from Solomon, Rivolo \& Barrett (1987) in Fig.\,\ref{larson}. The empirical relationship between the $M_{\rm vir}$ and $L_{\rm CO}$ follows from the predicted luminosity-linewidth and luminosity-size relationships of Larson (1981) following Maloney (1990). Observing Fig.\,\ref{larson}, we find that the bulk of the molecular clumps detected in MB\,-C follow roughly the relation predicted for Galactic molecular clouds, as, compared to the properties of molecular cloud cores in the Galaxy, they lie in a similar parameter space when accounting for median errors. They are marginally bound, and virialized, suggesting they are related to star-formation. 

The clumps in 30 Doradus and the outer Galaxy do not follow these relations. The molecular clumps in MB\,-C are regions of significantly higher gas densities within the low density Magellanic Bridge, and MB\,-C lies approximately at the boundary between the low and intermediate density regime for star formation (see Lehner 2002; Mizuno et al. 2006 for a discussion of gas densities in the Magellanic Bridge).


\subsection{Uncertainties on clump parameters}

 Uncertainties on the physical parameters of the clumps can be caused due to noise. A correction for a substantial sensitivity bias is made by extrapolating each parameter to find a value that we would expect to measure with perfect sensitivity (i.e. $\sigma_{\rm r.m.s.}$\,=\,0\,K). The extrapolation causes an enlargement effect on the parameters which can increase their uncertainties. On the other hand, the sizes and the velocity dispersion are corrected by the resolution bias, and such corrections are made using the quadratic subtraction of the two-dimensional beam and channel width, respectively. For those clumps which are not resolved by the beam size but have high signal-to-noise ($>$10), we have used the standard deconvolution method to estimate their radii assuming that the clump minor-axis is equal to the beam minor-axis, and deconvolving through the geometric mean. Such radii are considered as upper limits.


\subsection{Variations in the $X_{\rm CO}$ factor}

While CO and H$_2$ both trace the molecular component of the interstellar medium, they have rather different origins. H$_2$ forms on surfaces of dust grains, and can protect itself from ultraviolet (UV) radiation. CO requires much higher column densities to self-shield, and forms almost exclusively in the gas phase. The CO/H$_2$ ratio varies, and the conversion factor $X_{\textrm{CO}}$ from observed CO mass to total molecular mass (in H$_2$) is known to increase with decreasing metallicity, and has a non-linear dependence on luminosity (Bolatto, Wolfire \& Leroy 2013).

The clump mass, and surface density estimated previously are based on the virial theorem, therefore unaffected by a variation in the $X_{\rm{CO}}$ factor observed at lower metallicities. However, on comparing the total virial mass for each clump against the CO luminosity, to obtain an estimate of the conversion factor for molecular clumps in the Magellanic Bridge, we find that the mean value for these clumps determined by a least squares fit to the data is around 5.5.$\pm$4.6\,$\times$\,10$^{20}$\,cm$^{-2}$(K\,km\,s$^{-1}$)$^{-1}$, roughly similar to the Galactic value at these luminosities. The $X_{\rm CO}$ value is two times lower than the value of 1.4\,$\times$\,10$^{21}$\,cm$^{-2}$(K\,km\,s$^{-1}$)$^{-1}$ for the Magellanic Bridge reported by Mizuno et al. (2006) based on the same method, namely comparing the CO luminosities and virial mass for all the CO clouds in the SMC but at a  much larger scale size. The difference in the conversion factor is explained thus: the parsec scale CO clouds in the low metallicity environment of the Magellanic Bridge represent the dense region of the molecular H$_2$ cloud.  
Similar results have  been found in other low metallicity galaxies when the CO clouds are resolved at parsec scales (Rubio et al 2015, Schruba et al. 2017, Jameson et al. 2018, Salda{\~n}o et al. 2018, Valdivia-Mena et al. 2020). The CO clouds observed with ALMA show similar linewidths as the NANTEN single dish spectrum (the latter at spatial resolution of 156$\arcsec$). Taken together, this indicates the $X_{\textrm{CO}}$ factor tends towards the Galactic value when CO and the H$_2$ clouds are resolved to similar sizes (Rubio, Lequeux \& Boulanger 1993). To confirm this tendency, additional regions in the Magellanic Bridge have to be studied. A detailed study of the molecular clumps in the Magellanic Bridge is beyond the scope of this paper, but will be presented in a future study.

\section{Infrared SED fitting of young stellar objects}

Young, forming protostars are oft found fully or partially embedded in the clouds of gas and dust within which they form. The enshrouding dust renders most Galactic YSOs invisible at optical wavelengths. Infrared (IR) observations ($>$2\,$\mu$m) at wavelengths greater than average dust grain sizes help trace re-radiated energy from the stellar component of the protostar, as well as the surrounding circumstellar environment, offering an insight into their evolutionary stage. 

To interpret and classify YSOs, we match the observed spectral energy distribution (SED) against model predictions (from Robitaille et al. 2017). This additionally allows to gauge the potential range of physical parameters (principally the temperature and radius of the central source, and the surrounding disc/envelope component) that results in the observed flux distribution, enabling classification of the central source by estimating its evolutionary stage based on the dominant component.

\subsection{Data}

We compiled a list of archival infrared imaging by firstly searching for photometry in the {\it Spitzer} space telescope IRAC (Infrared Array Camera) mIR imaging at 3.6, 4.5, 5.8, 8.0$\mu$m, and MIPS (Multiband Imaging Photometer) 24$\mu$m archives covering the field-of-view of the ALMA observations. The data was taken as part of the {\it Spitzer} legacy program SAGE-SMC (Spitzer Survey of the Small Magellanic Cloud: Surveying the Agents of a Galaxy's Evolution; Meixner et al. 2006). A search across the region resulted in the retrieval of six point-like sources. The full width half maximum (FWHM) is 1.6$\arcsec$, 1.7$\arcsec$, 1.7$\arcsec$, 2$\arcsec$, and 5.9$\arcsec$ for the 3.6$\mu$m, 4.5$\mu$m, 5.8$\mu$m, 8.0$\mu$m, and 24$\mu$m images, respectively, which results in approximate resolution of 0.5--1.7\,pc\footnote{We cross-matched the {\it Spitzer} photometry to optical imaging (having FWHM of $\sim$0.5$\arcsec$) described in Section 4, and use the optical photometric ID to identify each source. The nomenclature for all sources throughout the paper follows from the optical photometry, which is tabulated in Table 3 and detailed in Section 4.}. {\it ri} photometry (described in Section 4) of optical counterparts, having FWHM of 0.5$\arcsec$ is included.

At the observed resolution of the mIR photometry, in addition to the primary point-like sources, it is probable that the FWHM includes contamination from diffuse or compact \ion{H}{II} regions, multiple sources, or small scale dust features which are unresolved. For this reason, we caution the reader that the final characterisation of the YSOs while accurate, might not provide a precise estimate of their properties. Moreover, long wavelength data ($\sim$70$\mu$m) which can be crucial to resolve the degeneracy between models although available (Seale et al. 2014), have much larger FWHM ($\gtrsim$5\,pc) at these distances, often including blends of multiple sources, and a contribution from the surrounding environment. Optical {\it ri} photometry is of higher angular resolution, and may exclude multiple sources that are captured due to the lower angular resolution in the mIR photometry. To account for this, optical photometry is only input as a lower limit in the SED fitting when the infrared sources are not resolved.

\subsection{YSO SED fitting}
To classify each source detected in the mIR, we construct SEDs for each of the six sources detected$^1$. The observed SEDs are fit against the Robitaille (2017) model grid of young stellar objects using the Robitalle et al. (2007) fitting algorithm. For the fitting, we fixed the distance to 59\,kpc, and adopted the Weingartner \& Draine (2001) extinction law for the SMC bar. The interstellar extinction range is between $A_V$ of 0.0 to 20\,mag, which is an additional component to the intrinsic dust caused by the circumstellar environment in the YSO model. We only fit sources with at least five flux measurements ($n$), providing a minimum constraint against overfitting. 

\begin{figure*}
        \centering
        \begin{subfigure}[b]{0.475\textwidth}
            \centering
            \includegraphics[width=\textwidth,page=1]{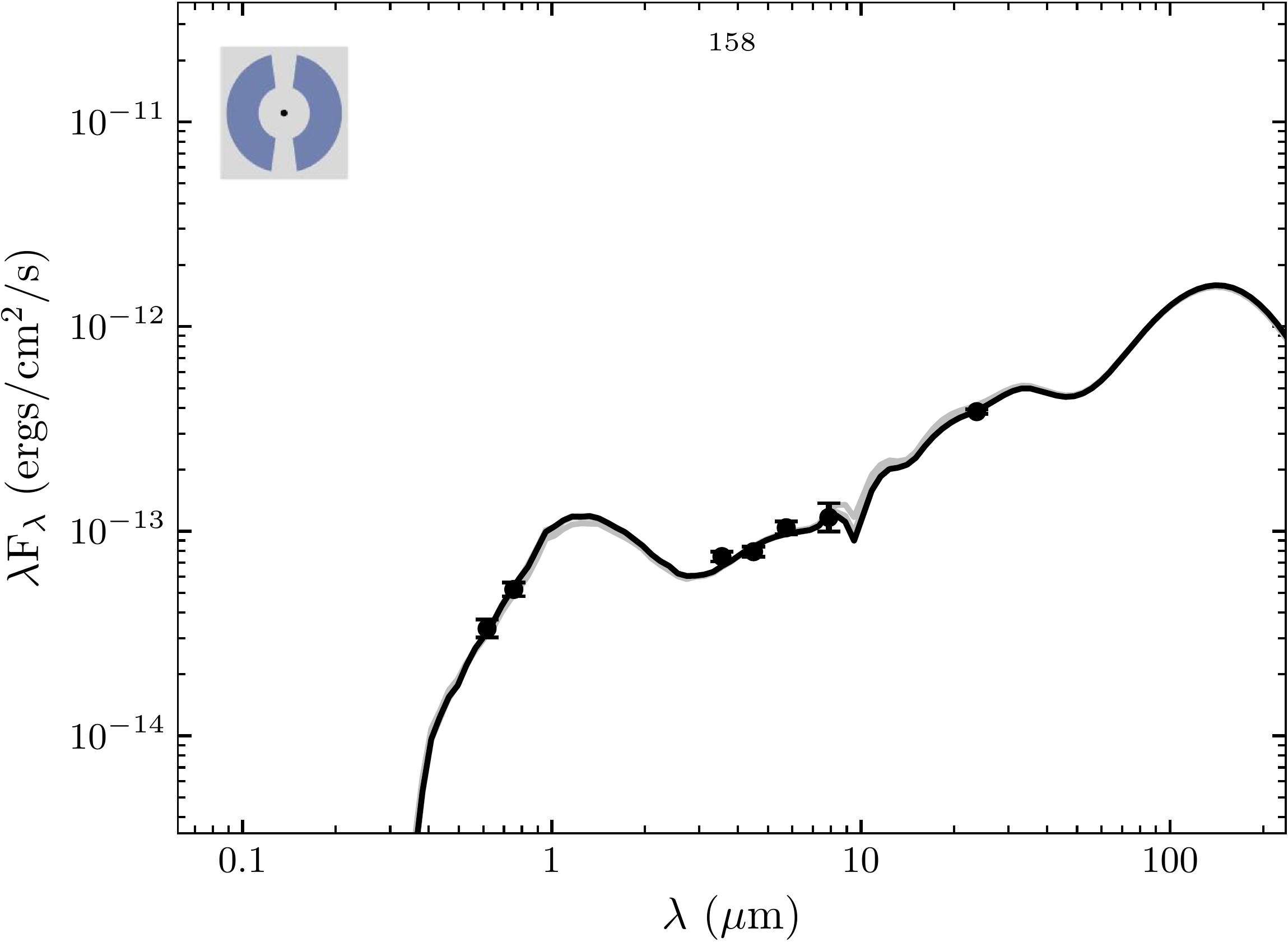}
            {{\small  }}    
            \label{fig:mean and std of net14}
        \end{subfigure}
        \hfill
        \begin{subfigure}[b]{0.475\textwidth}  
            \centering 
            \includegraphics[width=\textwidth,page=2]{wg8-crop.pdf}
            {{\small }}    
            \label{fig:mean and std of net24}
        \hfill
        \end{subfigure}
                \begin{subfigure}[b]{0.475\textwidth}
            \centering
            \includegraphics[width=\textwidth,page=3]{wg8-crop.pdf}
            {{\small  }}    
            \label{fig:mean and std of net14}
        \end{subfigure}
        \hfill
        \begin{subfigure}[b]{0.475\textwidth}  
            \centering 
            \includegraphics[width=\textwidth,page=4]{wg8-crop.pdf}
            {{\small }}    
            \label{fig:mean and std of net24}
        \hfill
        \end{subfigure}
        \begin{subfigure}[b]{0.475\textwidth}
            \centering
            \includegraphics[width=\textwidth,page=5]{wg8-crop.pdf}
            {{\small  }}    
            \label{fig:mean and std of net14}
        \end{subfigure}
        \hfill
        \begin{subfigure}[b]{0.475\textwidth}  
            \centering 
            \includegraphics[width=\textwidth,page=6]{wg8-crop.pdf}
            {{\small }}    
            \label{fig:mean and std of net24}
        \hfill
        \end{subfigure}
    \caption{SED of candidate YSOs having mIR photometry from 3.6--24$\mu$m fit against the Robitaille (2017) YSO models (Panels 1-6). The IDs are also indicated, along with the appropriate infographic taken from Robitaille (2017) for the best-fit stellar model in the upper left hand of each panel. The solid black line in each panel shows the best-fit YSO model, with filled circles flux representing values from mIR imaging. Triangles represent lower limits for fluxes from optical $r,i$, and downward triangles respresent upper limit for mIR 24$\mu$m photometry. Grey lines are remaining models which can also be fit within the best-fit criteria outlined in Section 3.2. The results of the fitting process are given in Table 2.}
    \label{fits}
\end{figure*}

Our resulting fits are shown in Fig.\,\ref{fits}. Models having passive discs, and in some cases power-law envelopes with cavities provide acceptable goodness-of-fit results, chosen here as $\chi^2$/$n\leq$3. To distinguish between models, we adopt the selection criteria laid out by Robitaille (2017), where the number of best fit models in each model set is calculated according to $\chi^2\leq\chi_{\rm{best}}+3$. Here, $\chi_{\rm{best}}^2$ is goodness-of-fit value for the best fitting model over all sets. We then identify the most probable set of models by finding the set which has the highest ratio of best fits to total number of models in the set. From the most probable model set, we select the best-fit model based on the $\chi^2$ value. As each set of models is a unique combination of different components, the most probable set of models will get a large number of well fit models, but a single low $\chi^2$ value in particular set of models may be caused due to overfitting, or simply coincidence. The most probable set of models, along with the name of the best fit model from that set, and the resulting central source parameters are given in Table\,2.

\subsection{Caveats of the SED models}
Numerous caveats exist when estimating YSO parameters based on SED models, which are outlined in Robitaille (2017). Briefly, the models only consider a singular source of emission, and cannot account for external radiation fields, polycyclic aromatic hydrocarbons (PAH) emission, or explicit accretion (i.e. they cannot model precisely UV or optical fluxes in accreting YSOs, but set a lower limit). Stellar parameters, such as mass or age are degenerate across models, and are determined by interpolating the model estimated central source temperature and radius over stellar isochrones and mass tracks. 


The best fit model is not intended to provide accurate parameters for each source, but is a crude guide to its nature, as the models have limited free parameters and are strongly degenerate (Offner et al. 2012). Moreover, ideal isolated cases are likely to be more accurate than in regions where angular resolution is an issue. In the latter scenario, modelling the SED of protostars suffers strongly from degeneracy and contamination from outflows and heating from nearby sources. Hence, we examine mIR imaging against higher resolution optical imaging (described in Section 4) shown in Fig.\,\ref{thumbnails} for each source to check for contamination.

\begin{table*}
	\centering
	\caption{Results of YSO model fitting to the IR SEDs, and the derived stellar parameters. }
	\label{tab:d}

\begin{adjustbox}{max width=\textwidth}
  \begin{threeparttable}
\begin{tabular}{lcccccccccccc}
\hline
  \multicolumn{1}{c}{ID} &
  \multicolumn{1}{c}{$n_{\rm pt}^a$} &
  \multicolumn{1}{c}{Model} &
  \multicolumn{1}{c}{Model} &
  \multicolumn{1}{c}{${\chi_{\rm best}}^{2^c}$} &
  \multicolumn{1}{c}{$n_{\rm fits}^d$} &
  \multicolumn{1}{c}{$R_{\ast}$} &
  \multicolumn{1}{c}{$T_{\ast}$} &
  \multicolumn{1}{c}{$A_V$} &
  \multicolumn{1}{c}{$L_{\ast}$} &
  \multicolumn{1}{c}{Mass} &
  \multicolumn{1}{c}{Log\,Age} &
    \multicolumn{1}{c}{YSO}\\
  \multicolumn{1}{c}{ } &
  \multicolumn{1}{c}{} &
  \multicolumn{1}{c}{Set$^b$} &
  \multicolumn{1}{c}{Name} &
  \multicolumn{1}{c}{} &
  \multicolumn{1}{c}{} &
  \multicolumn{1}{c}{($R_{\odot}$)} &
  \multicolumn{1}{c}{(K)} &
  \multicolumn{1}{c}{(mag)} &
  \multicolumn{1}{c}{($L_{\odot}$)} &   
  \multicolumn{1}{c}{($M_{\odot}$)} &
  \multicolumn{1}{c}{(yr)} &
    \multicolumn{1}{c}{Class}\\
\hline
\hline
  325 & 7 & sp-\,-h-i & Zdy7f7A5\_09 & 9.27 & 31 & 31.4$^e$ & 17686$^e$ & 4.73$^e$ & 4.9$^e$ & 23.7$^{f}$ & 3.9$^{f}$ & II/III \\
    &      &   (passive disc;      &               &       &   & (22.5--38.4) & (8715--23230) & (0.38--8.46) & (3.4--5.6) & (8.8--30.6)$^f$ & (4.1--6.8)$^f$ & \\
    & & variable $R_{\rm{inner}}$) & & & & & & & & & \\
              & & & & & & & & & & & \\
  158 & 7 & s-pbhmi & k42bZJxX\_02 & 11.26 & 8 & 9.4 & 9823 & 2.42$^e$ & 2.7 & 4.7 & 5.5 & I\\
      &      &   (Power-law envelope+cavity    &               &       &   &  &  & (2.06-2.45) &  &   &  &  \\
          & & +medium; variable $R_{\rm{inner}}$) & & & & & & & & & \\
                    & & & & & & & & & & & \\
  200 & 7 & s-pbhmi & jZVXXLnM\_08 & 13.45 & 2 & 19.8 & 6111 & 2.31$^e$ & 2.1 & 4.0 & 5.5 & I\\
      &      &    (Power-law envelope+cavity    &               &       &   & & & (2.27--2.33) & & & &  \\
          & & +medium; variable $R_{\rm{inner}}$) & & & & & & & & & \\
                        & & & & & & & & & & & \\
  206 & 5 & sp-\,-h-i & EQPlnTgq\_03 & 3.4 & 11 & 5.9$^e$ & 12469$^e$ & 1.78$^e$ & 2.9$^e$ & 5.0 & 5.5\\
      &      &   (passive disc;      &               &       &   & (5.8--6.3) & (10530-13340) & (0.3--2.2) & (2.6--3.0) & (4.4--5.2) & (5.5--5.6) & II/III\\
                & & variable $R_{\rm{inner}}$) & & & & & & & & & \\
                          & & & & & & & & & & & \\
  179 & 5 & sp-\,-h-i & rvs6hTE1\_02 & 1.46 & 1 & 21.3 & 14050 & 5.1 & 4.2 & 13.5 & 4.4 & II/III\\
      &      &   (passive disc;       &               &       &   &  &  & &  &  &  &  \\
          & & variable $R_{\rm{inner}}$) & & & & & & & & & \\
                    & & & & & & & & & & & \\
  269$^g$ & 7 & sp--h-i & HND0snQG\_02 & 45.9 & 1 & 29.3 & 5648 & 6.59 & 2.9 & 7.0 & 4.7 & II/III\\
    &      &   (passive disc;      &               &       &   &  &  & & \\
              & & variable $R_{\rm{inner}}$) & & & & & & & & & \\
                        & & & & & & & & & & & \\
\hline\end{tabular}
\begin{tablenotes} {
\item[]$^a$ is the number of valid data points used in the fitting. $^b$ is the best fitting set of models calculated based on the method outlined in Robitaille (2017). $^c$ is the $\chi^2$ for the best-fit model across all model sets. $^d$ is the number of models is the best-fit set of models with fitted models within the fit criteria. $^e$ The value is a weighted mean calculated following Carlson et al. (2011). $^f$ Some of the derived stellar temperatures and luminosities from the SED models fall outside the range of the stellar isochrones and mass tracks of Bressan et al. (2012). These values are not included in the results. $^g$ The $\chi^2$ value of the best fitting model does not fall within the fit criteria.}
\end{tablenotes}
  \end{threeparttable}
  \end{adjustbox}
\end{table*}

\subsection{YSO classifications based on infrared SED fitting}

\begin{figure*}
\center 
\includegraphics[width=17cm]{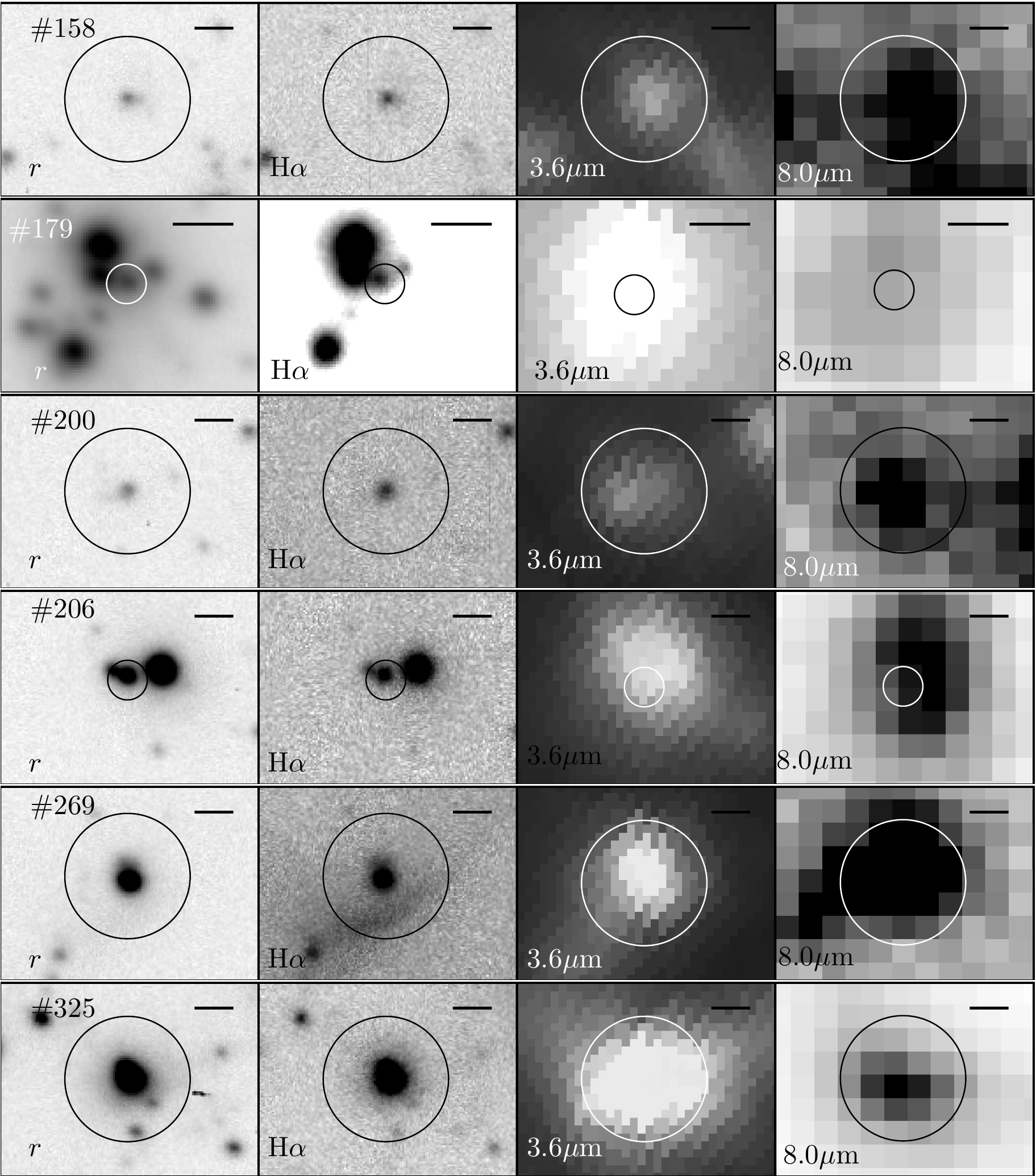}
\caption{Inverted linear greyscale thumbnails of all mIR sources in MB-C, along with their optical counterparts in $r$-band broadband, and H$\alpha$ narrowband imaging. The greyscale for \#179 is in logarithmic to highlight the multiple components. The panels from left to right show each individual source (top to bottom) in $r$, H$\alpha$, 3.6$\mu$m, and 8.0$\mu$m imaging respectively. All individual panels are labelled to the appropriate wavelength (lower left), and source (upper left). The scalebar in the upper left is of a length 2$\arcsec$, corresponding to $\sim$0.6\,pc at the assumed distance to MB\,-C. The solid circle marks an aperture of the source detected in the optical and infrared with a diameter of 4$\arcsec$, except in \#179 and \#206, where it marks an aperture of $\sim2\arcsec$. }
\label{thumbnails}
\end{figure*}

In this section, the SED fitting results, and the classification are discussed keeping in mind the caveats discussed previously. To adjudicate on the contribution from contaminants, we compare the {\it Spitzer} imaging to optical imaging (Fig.\,\ref{thumbnails}), with the latter having a mean resolution of $\sim$0.5$\arcsec$ (described in Section 4). The YSO model sets can be broadly related to the conventional YSO class schema (Lada 1987) based on the main contributing components for each model. For our analysis, each model set is classified based on the dominant component into sources with only envelopes (Class\,0; embedded sources), sources with both discs and envelopes, or envelopes with cavities (Class\,I), and sources with only discs (Class\,II), including sources with anaemic discs (Class\,II/III). Finally, we estimate the stellar parameters of each source based on the central source temperature and radius estimated from the best fitting model. Our results are compared to the previous classification of these sources carried out by Chen et al. (2014) in Appendix A1. The following description of each source follows the nomenclature described in Section 4$^1$.

Source \#158 appears isolated in both optical and infrared imaging (see Fig.\,\ref{thumbnails}). The resulting best-fit YSO model sets (Fig.\,\ref{fits}) are power-law envelopes with cavities, having emission from the ambient medium. The presence of an envelope in the best-fit model suggests the source is a young candidate Class\,0 YSO, as it is embedded inside of an envelope. Whereas the presence of a cavity suggests a more evolved source, suggesting a Class\,I stage. The best-fit model parameters  are given in Table 2. From these, we estimate the mass and age of the central source by interpolating against the model mass tracks and isochrones from Bressan et al. (2012) at $Z=1/5Z_{\odot}$ respectively, in the Hertzsprung-Russell diagram (HRD). The central source has a mass of 4.7\,$M_{\odot}$, and an age of 0.3\,Myr.

\#179 contains multiple near companions, and is faint in the optical imaging. The infrared photometry blends the multiple sources (up to 6 sources). The YSO models indicate a passive disc, i.e. a candidate Class\,II/III source. The contribution of unresolved stellar components to the SED likely results in an increase in flux at shorter wavelengths, and a later stage classification.

\#200 is well resolved in the optical and infrared imaging. The best fit set of models for \#200 consist of power-law envelopes with cavities. While the presence of envelope indicates a more embedded, younger stage of evolution, the cavity suggest a slightly more evolved central source. Based on this, we classify the source as a candidate Class\,I YSO. The resulting best fit parameters of the central source are a mass of 4\,$M_{\odot}$, and an age of 0.3\,Myr. 

\#206 has a brighter visual companion within 1$\arcsec$, which is not resolved in the {\it Spitzer} infrared imaging. The optical source also contains a faint companion within 0.5$\arcsec$. Further higher resolution imaging is required to resolve the source, and it is unclear if the {\it Spitzer} source corresponds to the optical source. The best fitting set of models contain a passive disc, indicating an evolved source (a Class\,II/III YSO candidate classification). The fit parameters, and classification is likely affected due to the unresolved component, as the YSO is fainter than the near companion. 

\#269 appears resolved in both the optical and infrared imaging, although the object appears extended in H$\alpha$. The best-fit set of YSO models suggest a passive disc, indicating a Class\,II/III YSO candidate classification. The best-fit model was chosen based on the lowest $\chi^2$ value amongst the model sets, and does not fall within the fit criterion. The stellar mass, and age is 7\,$M_{\odot}$, and $\lesssim$0.1\,Myr respectively.

\#325 is not resolved in the {\it Spitzer} infrared images, and the FWHM of the IR photometry contains emission from multiple sources, which appear resolved in the optical imaging. The best fitting set of models are comprised of passive discs. The dominant contribution of a passive/anaemic disc indicates a Class\,II/III YSO candidate classification. The source also displays PAH emission, as seen from the 4.5$\mu$m dip in the observed photometry (all other {\it Spitzer} bands are affected by PAH emission, hence their total flux will be larger than seen in 4.5$\mu$m resulting in a dip at that wavelength. Note that the models do not account for PAH emission). The presence of PAH emission suggests external heating from a nearby massive source (see Fig.\,\ref{thumbnails}), or possibly the central star itself. The classification, and the resulting central source parameters are affected by contribution from the unresolved sources. We note the similarity of the resulting set of fit models to \#179, and suggest that for both, the unresolved components significantly affect the resulting YSO classification.

\section{Accreting PMS stars from optical imaging}

As YSOs evolve from their natal cocoon, they collapse as pre-main sequence (PMS) stars surrounded by circumstellar discs, which are formed to conserve angular momentum as the protostar reduces in linear size by nearly two orders of magnitude. After collapse, these PMS stars have circumstellar discs which are truncated by their stellar magnetic field. Along these field lines mass is transferred. This accretion process results in gas flow along the magnetic field lines at this radius resulting in an energy release when impacting on the stellar surface. This energy from accretion heats and ionises the gas, resulting in strong H$\alpha$ emission (upwards of $-$10\AA\ in emission). This H$\alpha$ emission allows us to distinguish PMS stars (Class\,I-- Class\,III) undergoing accretion from the nascent stellar population which do not display H$\alpha$ emission at such magnitudes (Hartmann et al. 2016). Thus, H$\alpha$ studies of star-forming regions offer a snapshot of star formation in time for studying the properties of accreting PMS stars (e.g. Kalari 2019). In this section, we identify accreting PMS from deep optical photometry in MB-C based on their excess H$\alpha$ emission, and estimate their stellar properties (mass and age).

\subsection{Adaptive optics optical observations}

The MB-C region was observed using the SAM (SOAR Adaptive Optics module) imager mounted on the 4.1-m SOAR (Southern Astrophyiscal Research Observatory) telescope located at Cerro Pach{\'o}n, Chile in $r'i'$ filters along with a H$\alpha$ filter\footnote{The H$\alpha$ filter has a Full Width Half Maximum (FWHM) of 65\AA, and is centred on 6561\AA.} (key to identify H$\alpha$ emission line stars). We achieve ground-based angular resolutions around 0.5$\arcsec$ or better, essential to resolve individual stars at the distance to the Magellanic Clouds. Our observations are described in Kalari et al. (2018a). 

Briefly, the SAM imager covers a 3 arcmin sq. field on the sky at a pixel scale of 45.4\,{\it mas}. The MB\,-C region was captured with the imager centred on $\alpha\,=\,01^{h}56^{m}40^{s}9$, $\delta\,=\,-74\degr16\arcmin02\arcsec$. We took images in the $r'i'$ filters with individual exposure times of 10, 60, and 200s to cover a dynamic magnitude range without saturating the bright stars. In H$\alpha$, we took exposures of 60, and 200s. The airmass throughout our observations was between 1.39--1.42, with the final FWHM as measured from bright isolated stars using the {\textsf{IRAF}}{\footnote{{\textsf{IRAF}} is distributed by the National Optical Astronomy Observatory, which is operated by the Association of Universities for Research in Astronomy (AURA) under a cooperative agreement with the National Science Foundation.}} task {\it imexam} $\sim$ 0.4$\arcsec$-0.5$\arcsec$. Assuming a distance of 59\,kpc, the images are able to resolve stars further than 0.14\,pc apart.

To perform photometry on the images, we employed the {\textsc {STARFINDER}} package of Diolaiti, Bendinelli \& Bonaccini (2000). Following the same methodology of Kalari et al. (2018a), we detected sources, computed their photometry in the instrumental system, and calibrated our final magnitudes in the SDSS {\it ri} system, and tied our H$\alpha$ calibration to the $r$-band photometry. In total, 375 stars are detected in $r$, of which 120 stars have counterparts in both $i$ and H$\alpha$. These 120 stars form the final catalogue for our analysis. The formal uncertainties on our photometry are small and within 0.1\,mag for the faintest stars. We impose a cut-off in photometric random uncertainty of 0.15\,mag in each of our filters for the data used in further analysis. The region is not significantly clustered, and we checked for completeness using the {\textsf{IRAF}} task {\it addstar}. We find that the recovery fraction has no spatial dependence. The photometry reaches a depth of 3$\sigma$ at H$\alpha$ of 20.4\,mag. The resulting two-colour (TCD) and colour-magnitude diagrams (CMD) are shown in Figs.\,\ref{ccd} and \ref{cmdopt}.

\subsection{Identification of accreting PMS candidate stars}

The ($r-$H$\alpha$) colour is a measure of H$\alpha$ line strength relative to the $r$-band photospheric continuum. Main-sequence stars do not have H$\alpha$ in emission. Modelling their ($r-$H$\alpha$) colour at each spectral type allows for a template against which any colour excess due to H$\alpha$ emission can be measured from the observed ($r-$H$\alpha$) colour. The ($r-$H$\alpha$) colour excess (defined as ($r-$H$\alpha$)$_{\rm{excess}}$\,=\,($r-$H$\alpha$)$_{\rm{observed}}-$($r-$H$\alpha$)$_{\rm{model}})$ can be used to compute the H$\alpha$ equivalent width (EW$_{{\rm{H}}\alpha}$) (see De Marchi, Panagia \& Romaniello et al. 2010; Kalari 2019).  

We use the ($r-i$) colour as a  proxy for the spectral type. This results in the average error on the estimated spectral type classification to be between 2-3 spectral subclasses for the later spectral classes (F-K types), but of 4--5 subclasses for the early type stars, assuming that extinction is well determined. In the case of variable star-to-star extinction, spectroscopy is essential to estimate simultaneously extinction and spectral type. The effects of variable extinction on small spatial scales on our results is discussed in Section 4.5. The main-sequence colours at metallicity $Z\,=\,1/5\,Z_{\odot}$ for SOAR filters are taken from Kalari et al. (2018a).
 
\begin{figure}
\center 
\includegraphics[width=8cm]{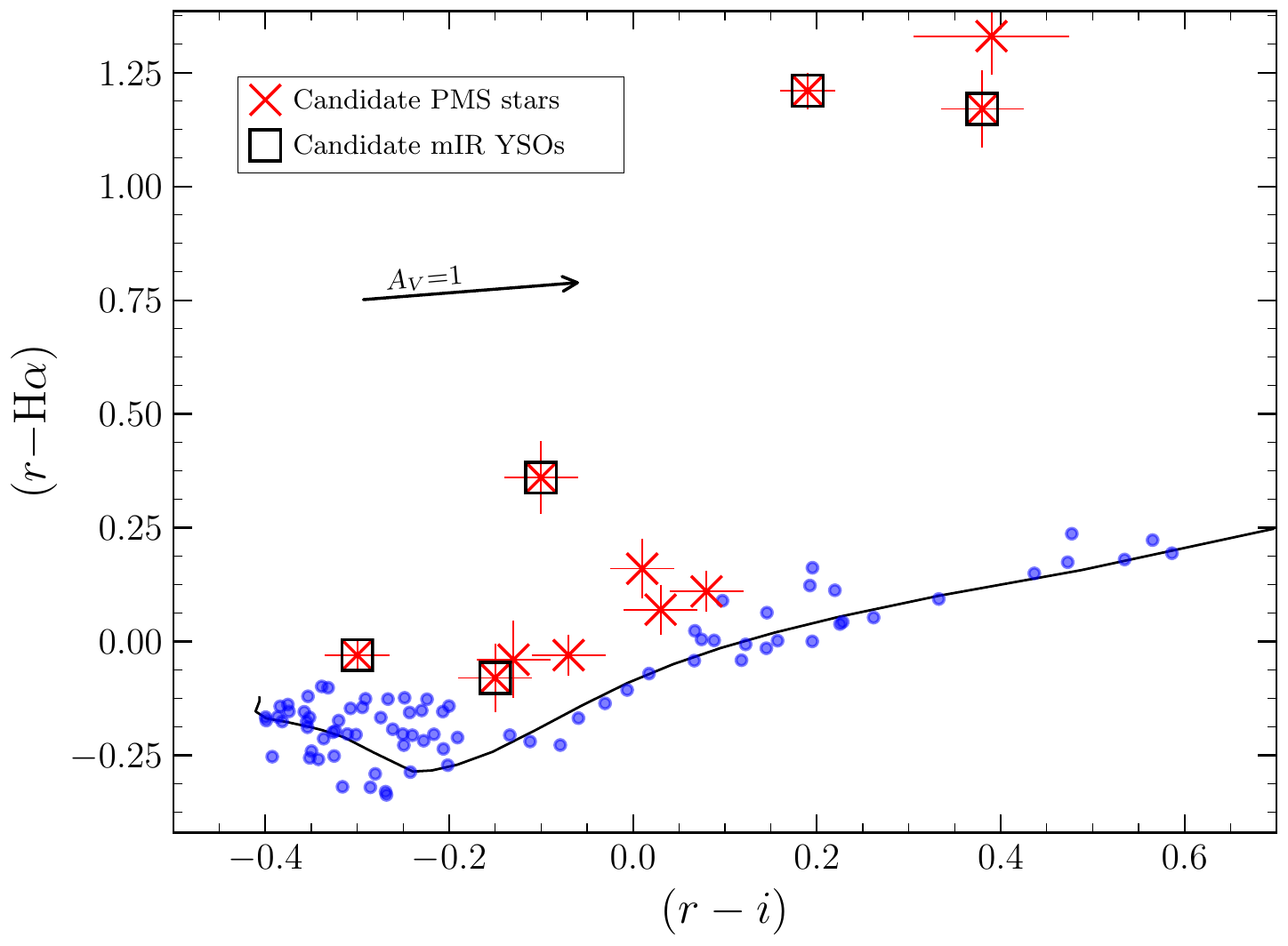}
\center
\caption{($r-$H$\alpha$) vs. ($r-i$) diagram. The solid line is the interpolated model track. Dots are all stars in MB-C having H$\alpha$ photometry, and crosses are stars identified as accreting PMS candidate stars based on their EW$_{\rm{H}\alpha}$. Squares mark stars having mIR counterparts. The position of the stars is corrected for extinction assuming $A_{V}$\,=\,0.5\,mag. The reddening vector for $A_{V}$\,=\,1 is also shown.}
\label{ccd}
\end{figure}

\begin{figure}
\center 
\includegraphics[width=8cm]{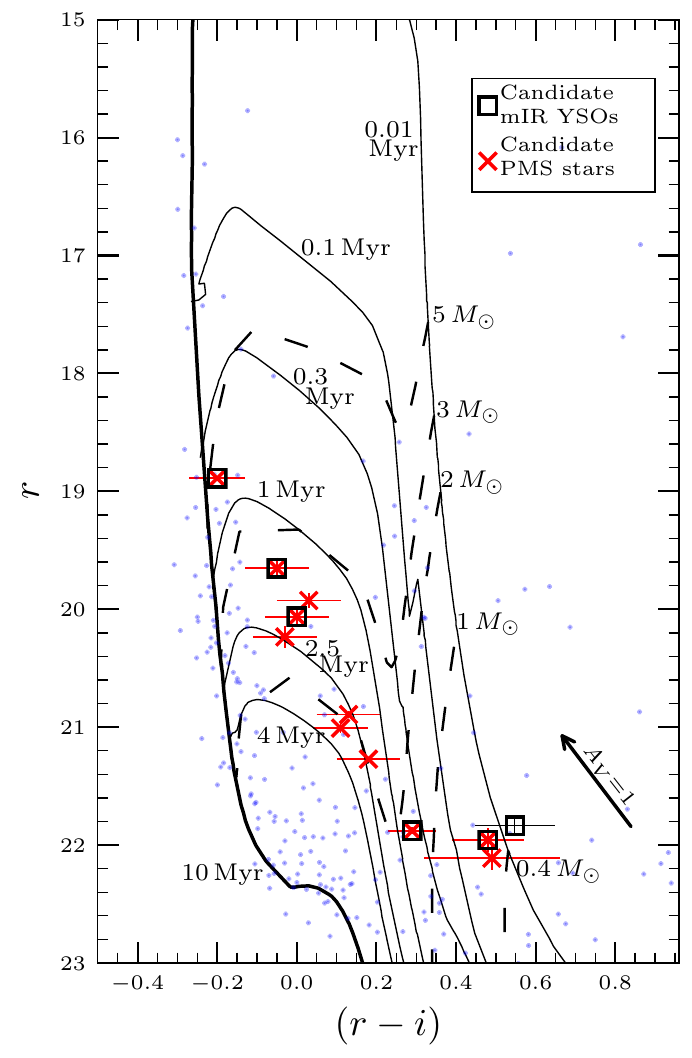}
\center
\caption{($r-i$) vs. $r$ CMD of all stars in MB-C shown as dots. The crosses are stars identified as optical candidate PMS candidates. Squares are stars identified as mIR candidate YSOs. Overlaid as solid lines are PMS isochrones from Bressan et al. (2012) of 0.01, 0.1, 0.3, 1, 2.5, and 10\,Myr. The dashed lines are stellar mass tracks of 0.4, 1, 2, 3, and 5\,$M_{\odot}$. The stellar tracks and isochrones are corrected for an extinction of $A_V$=0.5, and placed at a distance of 59\,kpc. The reddening vector of $A_V$=1 is also shown.  }
\label{cmdopt}
\end{figure}

In Fig.~\ref{ccd}, the ($r-$H$\alpha$) vs. ($r-i$) two-colour diagram is plotted. The solid line represents the interpolated model colours. We corrected for a mean absolute extinction $A_V$ of 0.5$\pm$0.07\,mag (where the error is the formal fit uncertainty) by fitting the model track to the main sequence locus of stars without any observable H$\alpha$ excess. This is lower than the mean value found for unblended sources by Chen et al. (2014).\footnote{We consider unblended sources as \#158, 200, 269 based on their {\it Spitzer} imaging in Fig.\,6. Chen et al. (2014) find a median $A_{V}$ of 2.5\,mag for these sources.} A detailed comparison between our values, and those in Chen et al. (2014) is found in Appendix A2. This suggests there likely exist locally enhanced extinction. However, the dust model used in the YSO models is not entirely appropriate for the Bridge environment. Hence, we refrain from simply adopting the extinction value found from fitting YSO models to the IR SEDs. We adopt a mean reddening of $A_V$=0.5\,mag as determined from the TCD, but also estimate shift in stellar parameters for values of $A_V$ of 1 and 2.5\,mag in Section 4.5.

At a given ($r-i$), the ($r-$H$\alpha$)$_{\rm{excess}}$ for each star is computed. The EW$_{\rm{H}\alpha}$ is given by
  \begin{equation}
    \ \rm{EW}_{\rm{H}\alpha}= \rm{W}\times[1-10^{(0.4\times(r-\rm{H}\alpha)_{\rm{excess}})}]\
  \end{equation}
following Kalari 2019. W is the rectangular bandwidth (similar in definition to the equivalent width of a line) of the H$\alpha$ filter. The photometric EW$_{\rm{H}\alpha}$ for all stars having $ri$H$\alpha$ magnitudes in our sample are measured. 

\begin{figure}
\center 
\includegraphics[width=6cm]{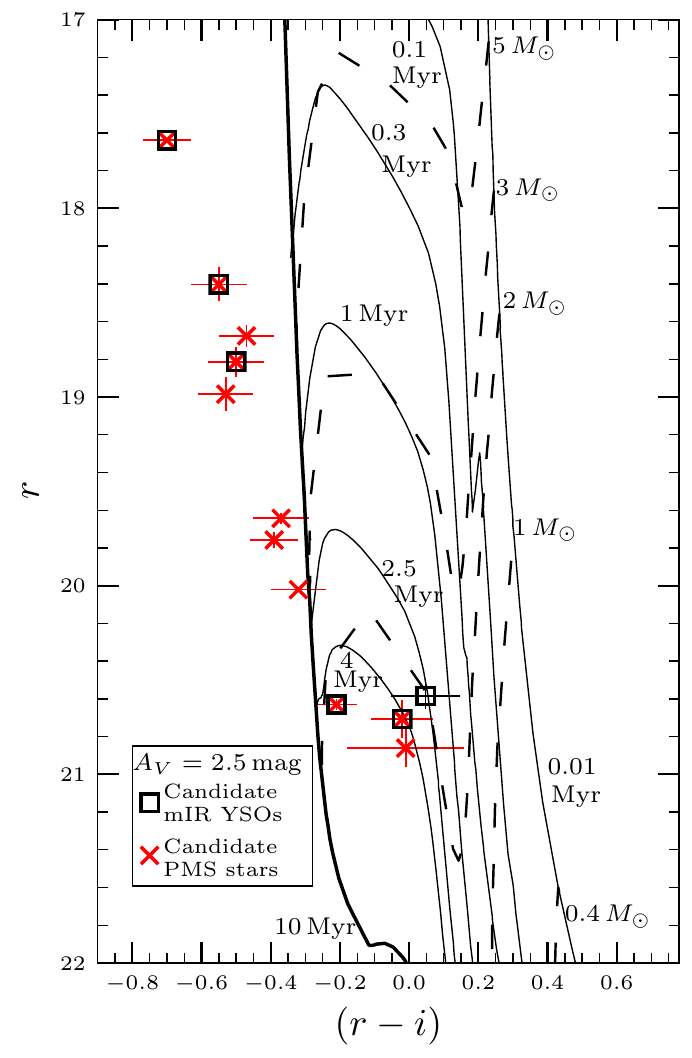}
\center
\caption{The CMD shown is same as Fig.\,\ref{cmdopt}, but with a mean extinction of $A_V$=2.5\,mag applied to the photometry. The stellar tracks, and isochrones are corrected by the distance modulus to MB-C. Symbols, and stellar models shown are same as Fig.\,\ref{cmdopt}. }
\label{cmdshift}
\end{figure}

 \begin{table*}
  	\centering
	\caption{Properties of candidate PMS candidates detected in SOAR optical $ri$H$\alpha$ imaging and their infrared counterparts}
	\label{tab:example_table}
 \begin{adjustbox}{max width=\textwidth}
 \begin{threeparttable}
    \begin{tabular}{llllllccccc}
\hline
  \multicolumn{1}{c}{ID} &
  \multicolumn{1}{c}{Right} &
  \multicolumn{1}{c}{Declination} &
  \multicolumn{1}{c}{$r$} &
  \multicolumn{1}{c}{($r-i$)} &
  \multicolumn{1}{c}{($r-$H$\alpha$)} &
  \multicolumn{1}{c}{Mass ($M_{\ast}$)} &
  \multicolumn{1}{c}{Age ($t_{\ast}$)} &
  \multicolumn{1}{c}{EW$_{\rm{H}\alpha}$} &
  \multicolumn{1}{c}{YSO} &
  \multicolumn{1}{c}{Fit} \\
    \multicolumn{1}{c}{} &
  \multicolumn{1}{c}{Ascension} &
  \multicolumn{1}{c}{ } &
  \multicolumn{1}{c}{(mag)} &
  \multicolumn{1}{c}{(mag)} &
  \multicolumn{1}{c}{(mag)} &
  \multicolumn{1}{c}{($M_{\odot}$)} &
  \multicolumn{1}{c}{(Myr)} &
  \multicolumn{1}{c}{(\AA)} &
  \multicolumn{1}{c}{class$^a$} &
  \multicolumn{1}{c}{Model$^b$} \\
\hline
\hline
  66 & $01^{h}56^{m}27^{s}36$ & $-$74\degr14\arcmin51\arcsec1 & 20.893$\pm$0.03 & 0.13$\pm$0.08 & 0.07$\pm$0.11 & 2.1$\substack{+0.2 \\ -0.4}$ & 2.6$\substack{+1.6 \\ -1.1}$ & $-$9.9$\pm$3.6 &  & \\
  82 & $01^{h}56^{m}32^{s}65$ & $-$74\degr16\arcmin52\arcsec9 & 21.008$\pm$0.04 & 0.11$\pm$0.07 & 0.16$\pm$0.13 & 1.9$\substack{+0.2 \\ -0.4}$ & 3.2$\substack{+1.6 \\ -0.9}$ & $-$16.2$\pm$5.6 &   &\\
  325 & $01^{h}56^{m}35^{s}62$ & $-$74\degr17\arcmin00\arcsec & 20.063$\pm$0.08 & 0.01$\pm$0.08 & 0.36$\pm$0.16 & 2.3$\substack{+0.1 \\ -0.2}$ & 1.5$\substack{+0.5 \\ -0.4}$ & $-$29.8$\pm$5.2 & II/III & sp-\,-h-i\\
  158 & $01^{h}56^{m}39^{s}03$ & $-$74\degr16\arcmin50\arcsec4 & 21.957$\pm$0.07 & 0.48$\pm$0.09 & 1.17$\pm$0.17 & 0.5$\substack{+0.1 \\ -0.3}$ & $\lesssim$0.1 & $-$44.8$\pm$16.1 & I & s-pbhmi\\
  200 & $01^{h}56^{m}40^{s}57$ & $-$74\degr16\arcmin53\arcsec4 & 21.88$\pm$0.03 & 0.29$\pm$0.06 & 1.21$\pm$0.08 & 1.6$\substack{+0.6 \\ -0.4}$ & 0.5$\substack{+0.3 \\ -1.0}$ & $-$47.6$\pm$18.7 & I & s-pbhmi\\
  88 & $01^{h}56^{m}42^{s}49$ & $-$74\degr17\arcmin27\arcsec & 21.271$\pm$0.01 & 0.18$\pm$0.08 & 0.11$\pm$0.09 & 2.1$\substack{+0.3 \\ -0.1}$ & 2.1$\substack{+1.6 \\ -2.0}$ & $-$10.3$\pm$4.7 &  & \\
  272 & $01^{h}56^{m}44^{s}42$ & $-$74\degr17\arcmin31\arcsec7 & 20.235$\pm$0.09 & $-$0.03$\pm$0.08 & $-$0.04$\pm$0.15 & 2.2$\substack{+0.1 \\ -0.2}$ & 2.4$\substack{+1.8 \\ -0.5}$ & $-$12.8$\pm$5.3 &  & \\
  20 & $01^{h}56^{m}45^{s}52$ & $-$74\degr15\arcmin01\arcsec8 & 19.926$\pm$0.06 & 0.03$\pm$0.07 & $-$0.03$\pm$0.09 & 2.5$\substack{+0.1 \\ -0.3}$ & 1.6$\substack{+0.5 \\ -0.4}$ & $-$9.4$\pm$6.0 &  & \\
  206 & $01^{h}56^{m}46^{s}35$ & $-$74\degr15\arcmin53\arcsec & 19.653$\pm$0.09 & $-$0.05$\pm$0.07 & $-$0.08$\pm$0.15 & 2.6$\substack{+0.2 \\ -0.1}$ & 1.5$\substack{+0.3 \\ -1.1}$ & $-$11.4$\pm$6.4 & II/III & sp-\,-h-i\\
  179 & $01^{h}56^{m}53^{s}78$ & $-$74\degr15\arcmin27\arcsec4 & 18.889$\pm$0.01 & $-$0.2$\pm$0.06 & $-$0.03$\pm$0.07 & 4.7$\substack{+1.2 \\ -1.0}$ & 
  1.0$\substack{+2.4 \\ -0.3}$ & $-$23.8$\pm$10.6 & II/III & sp-\,-h-i\\
  269 & $01^{h}57^{m}01^{s}93$ & $-$74\degr15\arcmin22\arcsec3 & 21.835$\pm$0.1 & 0.54$\pm$0.07 &  & 0.4$\substack{+0.2 \\ -0.1}$ & $\lesssim$0.1 &  & II/III & sp--h-i\\
  161 & $01^{h}57^{m}02^{s}76$ & $-$74\degr15\arcmin26\arcsec8 & 22.111$\pm$0.1 & 0.49$\pm$0.12 & 1.33$\pm$0.17 & 0.2$\substack{+0.0 \\ -0.8}$ & $\lesssim$0.1 & $-$48.4$\pm$11.8 &  & \\
\hline\end{tabular}
\begin{tablenotes}
{\item[]$^a$ Classifications based on Section 3.4. $^b$ Best-fit set of models from Robitaille (2017) following the $\chi^2$ criteria outlined in Section 3.2.}
\end{tablenotes}
\label{tableysos}
 \end{threeparttable}
\end{adjustbox}
\end{table*}

We homogeneously select accreting PMS candidate stars from the sample of stars having measured EW$_{\rm{H}\alpha}$ following the spectral type-EW$_{\rm{H}\alpha}$ criteria of Barrado y Navascu{\'e}s \& Mart{\'{\i}}n (2003). The median EW$_{\rm{H}\alpha}$ error due to photometric and reddening uncertainties is $\sim$\,10\,\AA\, so we empirically adjust the selection criteria by the same. We consider PMS candidates as those having spectral type later than A0 and earlier than K0 having EW$_{\rm{H}\alpha}$\,$<$\,$-15$\,\AA\ (i.e. beyond the criteria for a stellar wind from supergiants, due to a circumstellar discs from a Be star), or later than K0 and EW$_{\rm{H}\alpha}$\,$<$\,$-25$\,\AA\ (excluding chromospherically active late type foreground stars). The results of our selection procedure are shown in Fig.~\ref{ccd}. In total, eleven stars are selected on this basis. They are detailed in Table\,3.

Interestingly, we find that all PMS candidate stars identified in the infrared have probable counterparts in the optical (hence, optically visible candidate YSOs), with noted H$\alpha$ emission (except \#269, which displays extended emission in H$\alpha$). These results show similarity with those of Chen et al. (2014), who find that most candidate YSOs in the Bridge have optical counterparts, suggesting that in the Magellanic Bridge at lower metallicities, YSOs may have less dusty envelopes, and UV penetration is comparatively much deeper. If so, this may potentially allow us to study YSOs at earlier stages without the difficulties of high extinction faced in the Galaxy.

\subsection{Properties of PMS candidates}

For the PMS candidates we identified in the optical, we estimate the stellar properties (i.e. the stellar mass $M_{\ast}$, and stellar age $t_{\ast}$) by interpolating their position in the observed $r$ versus ($r-i$) CMD with respect to stellar tracks and isochrones in Fig.\,\ref{cmdopt}. The $r$-band magnitude includes an excess for H$\alpha$ emission line stars as it lies within the H$\alpha$-band, which is corrected for following the methodology of Kalari et al. (2015). For this purpose, we utilise the Bressan et al. (2012) single star isochrones and tracks as they cover the mass and age range of our sample at the appropriate metallicity (1/5\,$Z_{\odot}$; from Dufton et al. 2008). A comparison of the estimates of stellar mass and ages of PMS candidates in our mass range and metallicity value demonstrate that the Bressan et al. (2012) isochrones compare well with the isochrones from the PISA database (Tognelli, Prada Moroni \& Degl'Innocenti 2011) at ages later than 0.5\,Myr (see Kalari \& Vink 2015). The isochrones were reddened assuming the mean extinction of $A_V$=0.5. The errors on the derived mass and ages include the error on the reddening, and the photometric uncertainty. They are non-symmetric reflecting the non-uniform spacing of the stellar mass tracks and isochrones.

The ages of our sample range from $\lesssim$0.1 Myr to 3\,Myr. We find that the PMS candidates can be separated into two groups based on their ages. The first are six stars younger than 1.5\,Myr (\#158,\#179,\#200,\#206,\#269, and \#325), all of which have counterparts in the IR (note that \#269 is not identified as a H$\alpha$ excess emission star since it does not have acceptable H$\alpha$ photometry, rather its classification is solely based on its mIR photometry). Secondly, we detect five stars falling between ages 1.6--3\,Myr without IR counterparts (\#22, \#66,\#82,\#88,\#272). Given that our resolution in age (a median uncertainty around $\sim$1.5\,Myr) is smaller than the age differences between the two groups, it is not possible to arrive at firm conclusions regarding any differences between these populations. We do note that the older population has no mIR counterparts, and falls within a smaller mass range than the younger accretion PMS star population. A detailed comparison to the properties derived from SED fitting in this study, and in Chen et al. (2014) are presented in Appendix A2. We further explore spatial differences, to see if the older and younger populations show differences as seen in the Magellanic Clouds (e.g. De Marchi et al. 2010) in Section 5.

We find that our PMS candidates have masses between 0.2--4.7\,$M_{\odot}$. The PMS candidates with mIR YSO counterparts show a wide range of masses from 0.4\,$M_{\odot}$--4.7\,$M_{\odot}$. While the PMS candidates without mIR counterparts have masses between 0.2--2.5$\,M_{\odot}$, and are slightly older. However they are younger than the field stars detected towards the region (excluding for reddened foreground stars lying between 18$<r<$20). In Section 5, the properties of PMS candidate stars, and candidate mIR YSOs are compared with the spatial positions of the molecular clumps.

It is important to note that the precise ages (and masses) of individual stars derived using isochrones are highly uncertain, because the stellar properties may still be dependant on the accretion history and stellar birthline corrections, that are uncertain. However, the median age of a few stars is a more robust result, and provides a reliable indication on the relative age differences between regions (see Jefferies 2012 for an exhaustive discussion on this topic). For this reason, we employ the ages of the PMS candidates to study the star formation history of the region, but caution that the absolute value of stellar ages may not be precise. 


\subsection{Contaminants in the sample}

Multiple interlopers can mimic the position of H$\alpha$ excess stars in the TCD. The chosen H$\alpha$ EW criteria rules out the most common ones-- Be stars, and chromospherically active late type stars. However, it is possible other rarer early type emission line stars which have EW$<-15$\AA, such as reddened ($A_V>1$) B[e] stars, luminous blue variables (LBVs), or Wolf-Rayet stars (having extincted $r-i$ colours less than $-$0.2), might occupy positions similar to the bluest of our targets, given the direction of the reddening vector. It is impossible to completely rule out that such stars might have been mistakenly identified as accreting PMS stars without spectroscopy. However, LBVs, B[e] and Wolf-Rayet stars are extremely rare, with only one (Kalari et al. 2018b), four (Kraus 2019), and twelve (Massey, Olsen \& Parker 2003) known in the SMC respectively. Moreover, the latter two are found in extremely young high-mass star-forming regions. Therefore, while we cannot rule out the possibility of contamination by exotic massive emission line stars without spectroscopy, we consider this unlikely. 

At such distances, the angular resolution ($\sim$0.1\,pc) is not sufficient to rule out Ultra Compact \ion{H}{II} regions (UC\ion{H}{II}; Churchwell 2002). To measure the location of UC\ion{H}{II} in the TCD, we use the templates of Dopita et al. (2006) across a wide range of pressure and initial masses to simulate the model colours. We find that UC\ion{H}{II} models of Dopita et al. (2006), have $r-$H$\alpha$ colours $>$2 across a wide range of input parameters. Given the low extinction coefficient in this colour ($\lesssim$0.09; i.e. even at high or low extinction, the value of the $r-$H$\alpha$ colour for UC\ion{H}{II} regions remains the same), we consider confusion with UC\ion{H}{II} not important. However, only future long wavelength observations which can identify these regions conclusively, can resolve this issue (Hoare et al. 2007).    

Finally, \#179 is the only candidate PMS star having colours earlier than A0 spectral types (see De Marchi et al. 2010 for a more extensive discussion of the use of this cut-off to identify accreting PMS candidates in the Magellanic Clouds), and resembles a mid-late B spectral type based on it's ($r-i$) colour, and $r$-band magnitude. However, in the case it is an earlier spectral type, the presence of the observed dust might be due to a dusty compact H{\scriptsize II} region smaller than 0.1\,pc, which might also cause the observed H$\alpha$ excess emission. Further, the possibility also arises that the star could be a Be star, and the observed infrared excess is contributed in part from the surrounding stars (at least three nearby companions). Since without higher-angular resolution imaging, or emission line spectroscopy we cannot separate these scenarios, we refer to this caveat in future discussions of \#179.

\subsection{Uncertainties in the derived parameters}

The estimated parameters (mass, and age) have associated uncertainties (beyond those caused by photometric uncertainties) due to physical assumptions inherent in our analyses. These assumptions lead to non-systematic and systematic uncertainties in the estimated parameters. We consider here the result of systematic differences that may arise due to our assumptions in choice of mean extinction, and metallicity. The extinction, may vary from star to star (causing non-systematic uncertainties), but also the mean extinction may grossly under(or over) estimate the actual reddening towards MB\,-C. We consider that possibility here by estimating the stellar parameters assuming two different shifts in extinction. Additionally, the $Z$ towards MB\,-C is currently not known to the authors knowledge. The assumption of $Z$ similar to nearby massive stars (1/5$Z_{\odot}$; similar to the SMC main body following the study of hot stars in the Bridge by Dufton et al. 2008) may not be entirely appropriate. We consider what difference this assumption makes to the estimated stellar parameters by estimating the difference in stellar parameters if one chooses stellar models having metallicity of the LMC (1/2$Z_{\odot}$), and of the low-density region of the Magellanic Bridge (1/10$Z_{\odot}$; Rolleston et al. 1999).

Firstly, we recompute the stellar parameters (mass, age) after introducing mean shift in the assumed extinction by $A_V$=1, and 2.5\,mag. Considering a mean $A_V$ of 1\,mag, the median increase in stellar mass is 0.4\,$M_{\odot}$ (\#179 is not included as it falls outside the stellar mass tracks shifted to a distance of 59\,kpc at this extinction). The median stellar age increases by $\sim$ 0.3\,Myr. In the case of adopting a mean extinction of 2.5\,mag, eight stars out of our total sample fall outside the stellar mass tracks. The remaining stars (\#158, 161, 200, 269) have stellar masses between 2-3\,$M_{\odot}$. The median shift in stellar mass and age are 1.2\,$M_{\odot}$ and 4.5\,Myr respectively. The variation in stellar masses can be explained by looking at Fig.\,\ref{cmdshift}, where the stellar mass tracks follow the reddening vector for small values of extinction. Only stars approaching the main sequence show an increase in stellar mass under a moderate increase in extinction. The isochrones also follow the reddening vector for moderate values of extinction. However, at larger absolute extinctions, this trend is not replicated. The dereddened colours (following an extinction coefficient of 0.2 for the $r-i$ colour, this translates to a shift in colour of around $-$0.5) mostly fall outside the locus of stellar isochrones and mass tracks, making it necessary to extrapolate these values. The stars themselves would be considerably blue in the case of absolute extinctions around 2.5\,mag, lying beyond the zero-age main sequence in these cases. 

Secondly, we consider the shift in stellar parameters if the metallicity of the region is similar to the LMC (1/2$Z_{\odot}$), or to the low-density regime of the Magellanic Bridge (1/10$Z_{\odot}$) using the models of Bressan et al. (2012). The hydrogen and helium abundances were also scaled to $X=0.723,Y=0.267$ at 1/10$Z_{\odot}$, and $X=0.74,Y=0.256$ at 1/2$Z_{\odot}$, to reflect the abundances of the interstellar medium from which the stars form. The corresponding values of the hydrogen and helium fraction at 1/5$Z_{\odot}$ case were $X=0.746,Y=0.252$. In the 1/2$Z_{\odot}$ scenario, the mass of the PMS stars decreases by nearly a fifth of a solar mass, with a corresponding decrease in stellar age. In the latter scenario where the metallicity is similar to the low-density region of the Bridge, the masses of PMS stars are slightly larger in comparison to our results (by a value of 0.05\,$M_{\odot}$), and the ages by a larger shift of 0.3\,Myr. Thus, if the metallicity of MB\,-C was similar to the low density regions of the Bridge, and not similar to the SMC, the resulting shift in stellar parameters is considerably much smaller than if the metallicity of the region was underestimated. Thus, in the future if the metallicity of the region is revised to a value corresponding to the low density regions of the Magellanic Bridge (i.e. $\sim$1/10\,$Z_{\odot}$, the resulting stellar parameters would not change significantly.

 \begin{table}
	\centering
\caption{Summary of systematic uncertainties. Each column shows the mean shift in stellar mass and age for a simulated shift in extinction or metallicity in the adopted stellar isochrones and model tracks.}{
    \begin{tabular}{lccc}
\hline
 \multicolumn{1}{c}{Source} &
  \multicolumn{1}{c}{Simulated} &
  \multicolumn{1}{c}{Mass shift} &
  \multicolumn{1}{c}{Age shift} \\
    \multicolumn{1}{c}{ } &
  \multicolumn{1}{c}{shift} &
  \multicolumn{1}{c}{($M_{\odot}$)} &
  \multicolumn{1}{c}{(Myr)} \\
  \hline
\hline
   Extinction shift &  $A_V$=1\,mag & 0.4 & 0.3 \\
  Extinction shift & $A_V$=2.5\,mag & 1.2 & 4.5 \\
    Metallicity shift &   $Z=1/2\,Z_{\odot}$ & $-$0.2 & $-$0.45 \\
 Metallicity shift &    $Z=1/10\,Z_{\odot}$ & 0.05 & 0.3 \\
\hline\end{tabular}}
\end{table}

\section{Discussion}

\subsection{The star forming association MB\,-C}

In Fig.\,\ref{moneyplot} we show the spatial distribution of the identified candidate YSOs and PMS stars on an inverted greyscale H$\alpha$ image. The CO\,(1-0) contours are overlaid, which identify the star-forming molecular clumps described in Section 2. On examining the distribution, we find six sub-regions based on their location and proximity to individual molecular clouds. These regions are located within a large star-forming association, containing B stars which we term similar to the molecular cloud that encompasses it as MB\,-C. The overall properties resemble low density Galactic OB associations (e.g. de Zeeuw et al. 1999), with an approximate area of $\sim$50$\times$30\,pc, with only early B spectral type ionising stars (Chen et al. 2014). The sub-regions within the MB\,-C association are associated to molecular clumps, and in some cases ionising stars. We describe each sub-region below--

\begin{itemize}\setlength\itemsep{0.5em}

\item Region A: is a region with CO\,(1-0) emission located in the south-eastern region of MB-C. The total molecular clump mass is 296\,$M_{\odot}$ (clumps 5, and 8), with a median $\Sigma_{\rm H_2}$ of 72.5\,$M_{\odot}$\,pc$^{-2}$. It contains two candidate PMS stars identified on basis of their H$\alpha$ excess emission (\#88 and 272), none having mIR counterparts. The isochronal age of the candidate PMS stars are 2.1 and 2.4\,Myr, respectively. 
\begin{figure*}
\center 
\includegraphics[width=17cm]{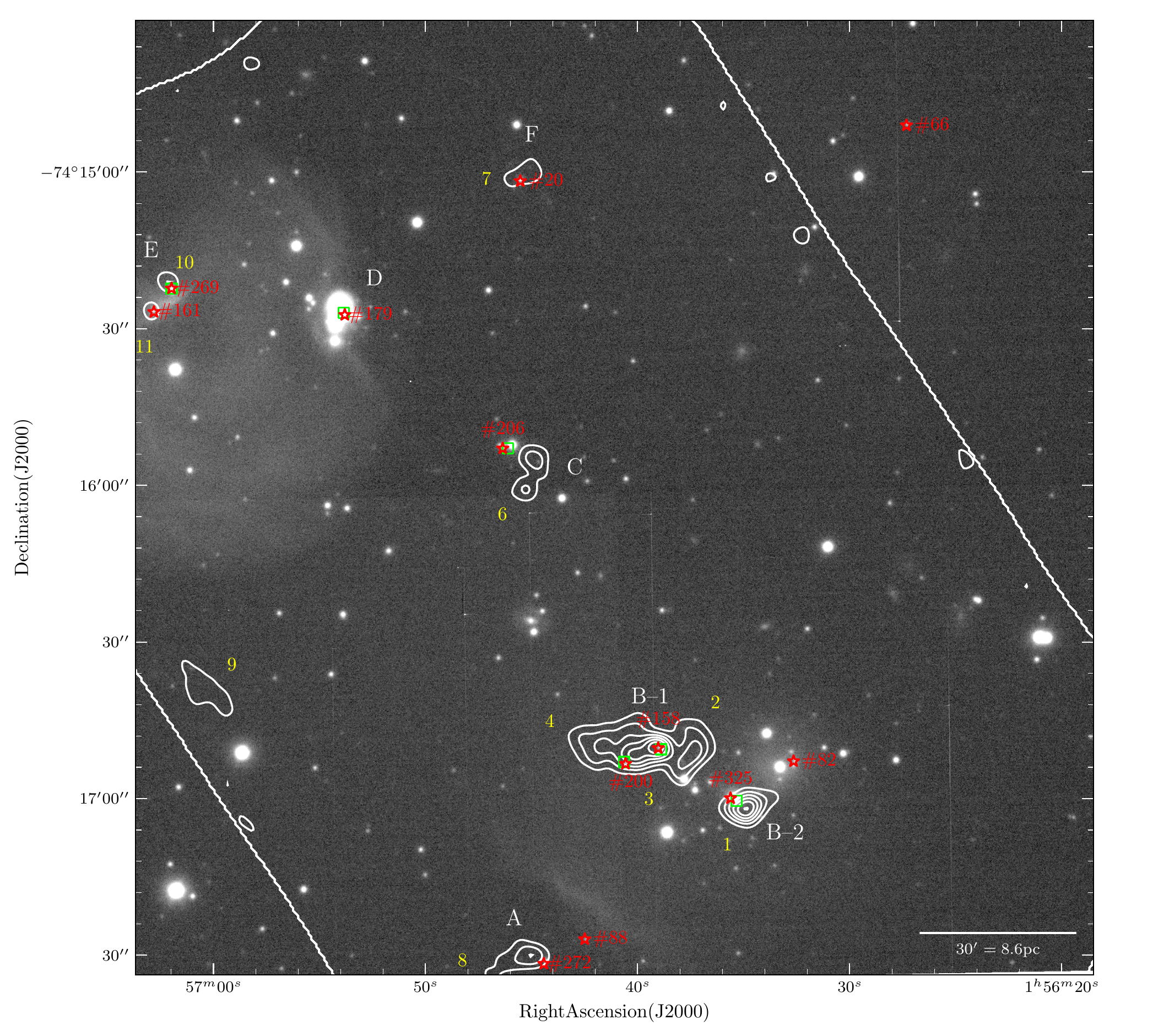}
\caption{The inverted H$\alpha$ greyscale image of MB-C. Each individual sub-region is labelled. The positions of the optically identified PMS candidates are shown as asterisk's. Contours represent the CO\,(1-0) identified star-forming clumps. mIR YSOs are marked as squares. Approximate molecular cloud positions are labelled in yellow following Table 1, and the PMS candidate stars identified in Table 3 are labelled in red. North is up and east is to the left, with a scalebar given in the lower right corner. }
\label{moneyplot}
\end{figure*} 

\item Region B: is the most interesting and largest sub-region. It can be further split into two, B-1 (clumps 2, 3, and 4) and B-2 based on the molecular clumps (clump 1) in sub-region B. PMS stars having mIR counterparts are located at the centre of both the molecular clumps, with B-1 having two (\#158 and 200) and B-2 one (\#325). The total molecular mass is 1747\,$M_{\odot}$, with the median $\Sigma_{\rm{H}_2}$ of 90\,$M_{\odot}$\,pc$^{-2}$. The association of YSOs likely bearing envelopes located at the centre of the molecular clumps in B-1 suggests that the region is extremely young compared to Region A. The median age of the PMS candidates shows a difference, with B-1 having PMS candidates less than 0.5\,Myr and B-2 having a PMS candidate of 1.5\,Myr (\#325) that shows evidence in its mIR SED for a passive disc, and also PAH emission. This suggests that there is active UV ionisation from the massive early B-type stars located in this region (see Chen et al. 2014). An older PMS candidate (\#82) with an isochronal age of 3.2\,Myr near B-2 is also detected away from the molecular clump. The region as whole exhibits H$\alpha$ nebulosity and is also bright in the mIR. The external stellar ionisation is responsible for the \ion{H}{II} region, and possibly also the faint \ion{H}{II} region in Region A given the latter's lack of ionising stars.

\item Region C: is a molecular clump (clump 6) containing one PMS star (\#206) having an isochronal age of 1.5\,Myr, with a nearby mIR YSO having a passive disc (Class\,II/III). The region is not H$\alpha$ bright. The total molecular mass is 566\,$M_{\odot}$ with a $\Sigma_{\rm{H}_2}$ of 80\,$M_{\odot}$\,pc$^{-2}$.

\item Region D: Is a bright \ion{H}{II} region ionised by two B1-B2V spectral type stars (Chen et al. 2014). The region contains multiple sources which are not resolved clearly in the mIR. No molecular clumps are detected towards this region. One PMS candidate (\#179) with an age of 1\,Myr is detected. It is the most massive PMS candidate having a stellar mass of 4.7\,$M_{\odot}$. However, it may not be resolved in optical and infrared imaging, and may be an early-type main-sequence star (see Section 4.4). Given the young age of the PMS candidate, but the lack of CO\,(1-0), we tentatively suggest that either the ionising stars may have blown away the natal molecular cloud giving rise to the bright \ion{H}{II} region observed, or higher resolution imaging is required to truly disentangle the YSOs from the stellar population.

\item Region E: contains two unresolved molecular clumps (clumps 10, and 11) having a total mass of 576\,$M_{\odot}$, and $\Sigma_{\rm{H}_2}$ of 116\,$M_{\odot}$\,pc$^{-2}$. One clump contains a mIR YSO (\#269, that is extended in H$\alpha$ counterpart), and one contains PMS candidate identified based on H$\alpha$ excess emission (\#161), without a mIR counterpart. The region is bright in both H$\alpha$ and the mIR. The age of the PMS stars is less than 0.1\,Myr.

\item Region F: Contains a 1.6\,Myr PMS candidate (\#20) located in the centre of a molecular clump (clump 7). The clump has a mass of 34\,$M_{\odot}$ and density of 17\,$M_{\odot}$\,pc$^{-2}$. The molecular clump deviates from the virial equilbrium relation, and caution is exercised on the precise mass and surface density values.

A single PMS star (\#66) is located outside the ALMA field-of-view, and we do not consider further in our work regarding the structure of the region. 

\end{itemize}

We find that the spatial location of forming stars is coincident with the molecular clumps indicating extremely young ages. There is no relation between the surface density of star-forming clump and the total number of YSOs, with YSOs forming from molecular clumps having surface densities greater than $\sim$17--200\,$M_{\odot}$\,pc$^{-2}$. This includes seven clouds whose surface densities are smaller than the commonly adopted Galactic threshold of $\sim$120\,$M_{\odot}$\,pc$^{-2}$ (Lada et al. 2010; Heiderman et al. 2010). Interestingly, we find that molecular clouds having lower surface densities (17--120\,$M_{\odot}$\,pc$^{-2}$; with a median surface density of 67\,$M_{\odot}$\,pc$^{-2}$) are associated to the older PMS stars ($>$1.5\,Myr). While, younger PMS stars ($<$1.5\,Myr) are found near molecular clouds having a higher median surface density of 93\,$M_{\odot}$\,pc$^{-2}$ (with a range between 57--200\,$M_{\odot}$\,pc$^{-2}$).

Of course, the dense gas in the current epoch is not the gas from which these stars formed, nor is it currently forming stars. But, under a simplistic steady state assumption, our results do not favour the surface density thresholds required for star formation commonly adopted in Galactic studies that assume a certain magnetic field strength (Heidermann et al. 2010). However, low-density gas in the Magellanic Bridge has been evolving into denser molecular gas, from which stars form, and the lower densities of the molecular gas in this region may signal the exhausting of the latter. 

The density of YSOs and PMS stars in the sub-regions range in a continuum from moderately clustered to relative isolation, reminiscent of star formation in the Milky Way (e.g. Bressert et al. 2010). However, low number statistics precludes us from defining a surface density for each sub-region. Our results suggest that star formation observed in MB\,-C is underway in localised regions depending on the density inhomogeneities within an initial cloud, and is intimately linked with the presence of dense molecular gas.

\subsection{Molecular gas structure in the Magellanic Bridge}

In Section 2, we have investigated the properties of the molecular clumps in MB\,-C. The clumps were gravitationally stable, and very similar to clumps related to star-formation in the Milky Way. Their densities, radii, masses, and linewidths do not differentiate them from clumps previously identified in the intermediate to low-density regimes of star formation in the Milky Way (Kennicutt \& Evans 2012). The clumps are related to star formation, as seen by their proximity to YSOs and PMS stars. The properties of YSOs and PMS stars show no remarkable differences compared to the Galaxy on account of their metallicities, except lower extinctions towards them. Overall, this suggests that even though the CO has lower filling factors in the Magellanic Bridge than found in solar-metallicity molecular clouds, the actual star formation process proceeds similarly once cold molecular gas is formed, and differences may arise due to the conversion of atomic to molecular gas. The lower dust-to-gas ratio results in lower extinction towards the line of sight of protostars, which affects similarly the $X_{\rm CO}$ factor, but the formation of molecular gas is the unifying factor between star formation. The effect of the lower self-shielding by CO on its durability is unclear, since the incident radiation field in MB\,-C is not particularly strong only containing early B-type stars. 

Recent observations from the {\it Herschel} space telescope at far infrared wavelengths have suggested a common unifying factor between star formation is the formation of filamentary structure in the cold interstellar medium (Andr{\'e} 2017), ubiquitous in star-forming regions. Unfortunately, the FWHM of {\it Herschel} imaging ($\sim$ 5--10\,pc) at the distance to MB\,-C is more than an order of magnitude larger than the average width of filaments detected in the Milky Way ($\sim$0.1\,pc), therefore it is currently not possible to identify filaments in the Magellanic Bridge.  

\begin{figure*}
\center 
\includegraphics[width=17cm]{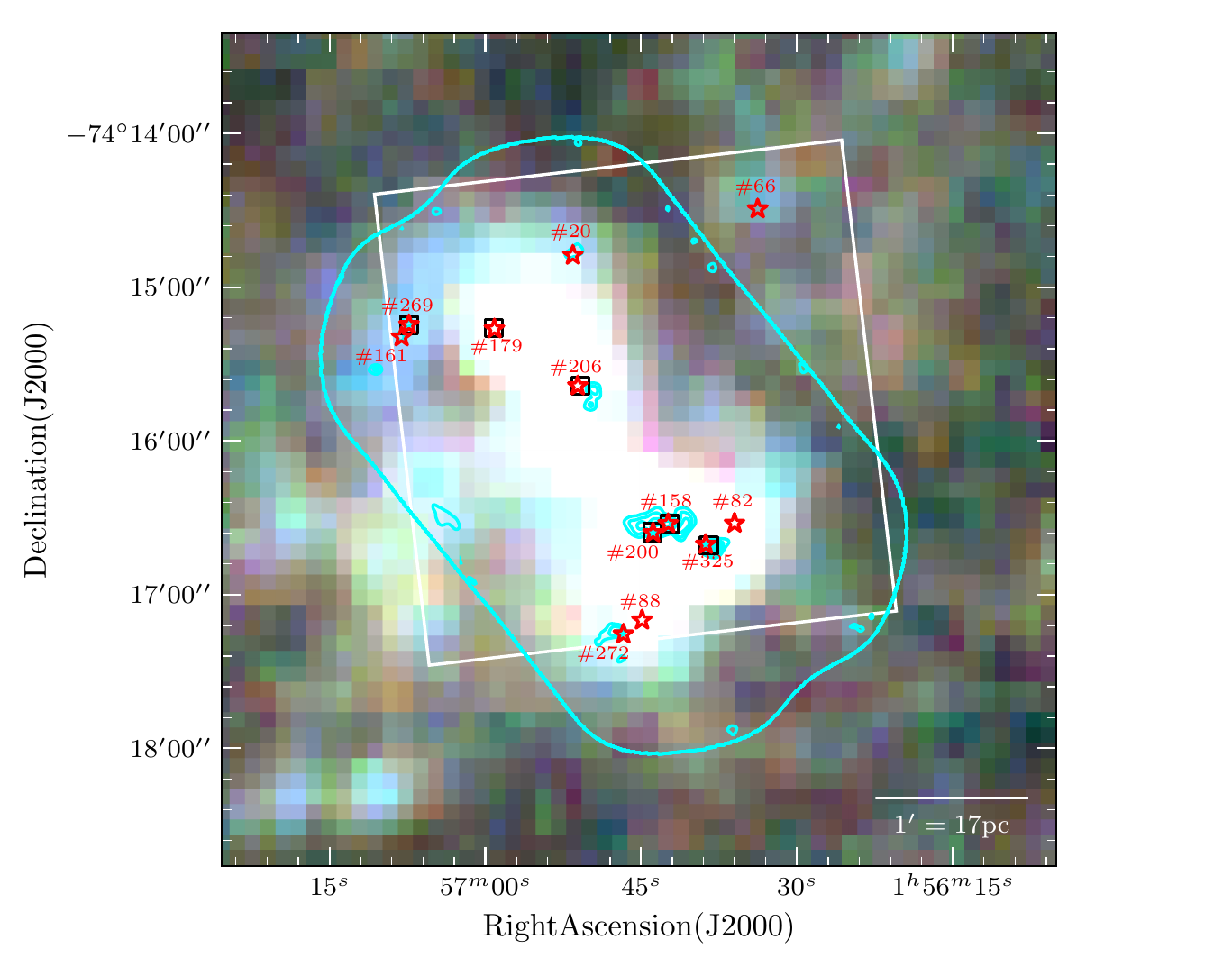}
\caption{500$\mu$m/350$\mu$m/250$\mu$m $r/g/b$ image of MB-C highlighting the filamentary structure. North is up and east is to the left. A scalebar is shown to the lower right. Overlaid are the FoV and CO(1-0) contours from the ALMA image in cyan, and the FoV of the SOAR observations in white. The filamentary structure running along the length of the region is seen. The positions of PMS candidates are shown as asterisk's, and the mIR identified YSOs are marked as squares. The YSOs all lie along the length of the filamentary structure. PMS candidate stars identified in Table 3 are labelled in red.}
\label{filament}
\end{figure*}

Instead, we simply overplot YSOs and PMS candidates on the 500$\mu$m/350$\mu$m/250$\mu$m {\it Herschel} {\it rgb} image of MB\,-C (Seale et al. 2014) in Fig.\,\ref{filament}. A clear, large-scale filamentary structure is observed, spanning the width of the association. The structure is approximately $\sim$30\,pc in length, and has a width corresponding to the resolution of the image ($\sim$5\,pc). The length of the major axis is comparable to some of the large filamentary structures found in the Milky Way (Hill et al. 2011). It is possible that smaller filamentary structures, with FWHM similar to those observed in the Milky Way can be identified with higher angular resolution images. The structure is related to star formation, as the youngest forming stars (see Section 5.3) fall along its length. The structure also contains the dense molecular clumps identified in CO\,(1-0). To form protostars, most filaments are considered to be supercritical, which is not the case for the observed molecular clumps. Supercritical filaments are unstable to gravitational collapse, and collapse to dense clumps or cores along the major axis. The location of the YSOs and PMS candidates, and dense molecular clumps supports this picture.

\subsection{Star formation in space and time in MB\,-C}

While the absolute ages of the PMS candidates identified in MB\,-C are not precise, their relative ages can be considered for further analysis, and are commonly viewed as accurate enough to search for differences in ages of populations in regions (see Jeffries 2012). This analysis is underpinned by the assumption that the distribution of stellar ages, and any differences in the ages of populations is caused principally due to the star formation history of a particular region (discounting the uncertainties on the ages). Younger PMS candidates, have not had much time to travel from their natal sites, and would be found closer to the location of their formation, and more clustered. Conversely, older PMS candidates in this scenario would have time to dynamically evolve and move further away from their location and therefore are to be found more spread apart. Age differences significantly larger than our mean errors ($\sim$1-1.5\,Myr) might be caused due to this dynamical evolution.

\begin{figure*}
\center 
\includegraphics[width=17cm]{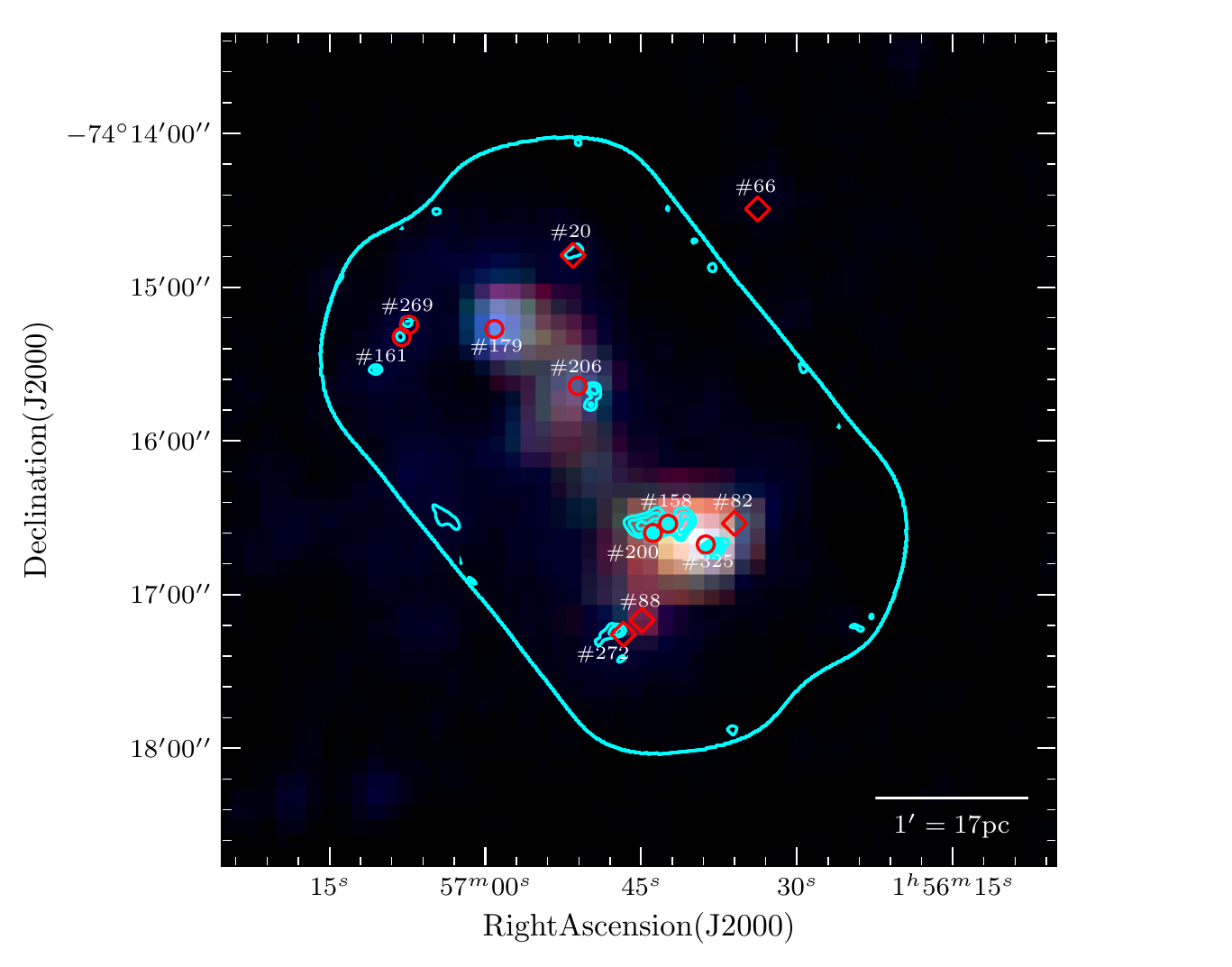}
\caption{500$\mu$m/350$\mu$m/250$\mu$m $r/g/b$ image of MB-C. North is up and east is to the left. A scalebar is shown to the lower right. Overlaid are the ALMA contours and FoV (solid line). The young PMS candidates ($<$1.5\,Myr) are marked by circles, while older PMS candidates ($>$1.5\,Myr) are shown as squares. PMS candidate stars identified in Table 3 are labelled.}
\label{ages}
\end{figure*}

In Fig.\,\ref{ages}, we plot the location of all PMS candidates having ages less than, and more than 1.5\,Myr
respectively (for reference, this is approximately the transition age between Class\,II/III intermediate mass YSOs; Mamajek 2009). The locations of PMS candidates are overlaid on the {\it Herschel} {\it rgb} (Seale et al. 2014) image, highlighting the filamentary structure of the molecular cloud (see section 5.2). Also shown are the CO\,(1-0) contours demarking the molecular cloud boundaries. A striking feature of this plot is the location of the young, and old PMS candidates. Younger PMS candidates, having ages less than 1.5\,Myr are distributed almost exclusively along the length of this filamentary structure, and older ones are more scattered from the structure. These young PMS candidates are found amongst the densest molecular cloud clumps. The median age of the PMS candidate stars is $\sim$0.5\,Myr, with all but one (\#161) being associated to mIR bright sources. Note that the Class\,II/III mIR detected YSOs \#179, and \#325 are located at the ends of the filamentary structure, while the major length of the filament harbours mostly PMS candidates identified as Class\,I YSOs using SED fits. 

Meanwhile, the older PMS candidates are distributed in the region away from the major axis of the structure, towards the sub-region A, and F. These regions are also associated to the molecular clumps having lower surface densities ($\sim$20--50\,$M_{\odot}$\,pc$^{-2}$; except for clump 5 in sub-region A). The median age of the PMS candidate stars in this region is also higher at $\sim$2.5\,Myr. Given that the median age uncertainty for the sample is $\sim$1.5\,Myr, the two populations are separated in age. The location of the two populations, young, and old, along, and away from the filament respectively, and the relation to the molecular cloud clumps and their properties, suggests that we are observing spatial morphologies of young stars, related in a very similar fashion to filaments and molecular clouds in the Milky Way. Although the exact details regarding star formation and filamentary structures are debated (Myers 2017), it is clear that star formation in MB\,-C displays similar morphologies to those found elsewhere in the Milky Way.

A major caveat to note here is that the age uncertainty in our study is not sufficient to disentangle whether star formation in MB\,-C has happened in ``bursts'' with the younger and older populations having formed separately (as was recently found by Beccari et al. 2017), or has proceeded slowly over time. The observed differences between the median age of both the young and old populations is around 2\,Myr, which is roughly similar to the median age uncertainty of our analysis, not allowing us to discern the burst scenario from the continuous star formation one. Support to our analysis of differences between the ages of the PMS populations is provided by detecting nearly all of the young PMS candidates are mIR bright sources, which are considered to be younger than solely optically visible accreting YSOs. We therefore can only conclude with confidence that most of the younger population of PMS candidates falls along the length of the visible filamentary structure, while the older population are more distributed hinting at dynamical evolution within the region. Overall, the association contains stars forming currently at ages $\lesssim0.1$--3\,Myr, with the younger PMS candidates located along the filament and the older PMS population more distributed.

Finally, we note that there exists other regions of higher \ion{H}{I} concentration in the Magellanic Bridge (see Fig.\,1), without known star formation. Literature searches for molecular material, or YSOs in the Magellanic Bridge have been restricted to R.A. less than 3$^h$ (see Section 1 for a brief review). There exist regions with known \ion{H}{I} concentrations, and blue main-sequence stars at R.A. between 3$^h$ and 4$^h$30$^m$ (see Irwin, Demers \& Kunkel 1990). It would be interesting to see with high-angular resolution and sensitivity CO observations, if molecular gas exists there, and with deeper IR observations if stars are also formed. The absence of both, or one may simulate interesting comparisons to MB\,-C.

\section{Summary}

A multi-wavelength analysis of the star-formation region MB-C in the Magellanic Bridge is presented using CO\,(1-0), infrared (3.6--24$\mu$m), and optical observations. The metallicity of early-type stars surrounding the region is 1/5\,$Z_{\odot}$. MB\,-C lies in a region of higher \ion{H}{I} compared to the low-density Bridge, with average gas surface densities larger than 10$^{21}$\,cm$^{-2}$. From resolved observations of the molecular clouds (at angular scales of 0.9\,pc), mIR ($\sim$0.5\,pc), optical ($\sim$0.15\,pc), we are able to study star formation in the region. 

From CO\,(1-0) observations obtained at a parsec size scale with ALMA, we detect twelve molecular clumps in large molecular cloud. The surface densities of these clumps range between 12--200 $M_{\odot}$pc$^{-2}$, comparable to Galactic star-forming clumps (e.g. Heiderman et al. 2010; Gutermuth et al. 2011; Heyer et al. 2016). From the analysis of the observed properties of the molecular clumps, we find that they are around sizes of 1\,pc, and are in gravitational equilibrium. 

From mIR imaging, we identified six YSOs detected based on their SEDs, which were previously identified in Chen et al. (2014). Fitting YSO models to the SED of the protostars, suggests they are young, Class\,I--III sources. Three protostars, whose SEDs may be contaminated due to the low angular resolution of the mIR imaging, were identified as Class\,II/III sources. The analysis suffers from effects of low angular resolution, and other limitations due to the YSO models themselves. 

We identified eleven H$\alpha$ excess sources from the optical $ri$ broadband, and H$\alpha$ narrowband imaging. Our results suggested a mean absolute visual extinction towards the optically visible population of 0.5\,mag. They were classified as PMS candidate stars currently undergoing accretion. Five of the stars identified have mIR counterparts, which were previously classified as YSOs. The PMS candidates fall into two broad age groups, young with median ages $\sim$0.5\,Myr, and older having median ages $\sim$2.5\,Myr. The mass range of our sample is $\sim$ 0.2--4.7$M_{\odot}$, suggesting we have identified accreting low-mass PMS candidates similar to those commonly found in such analyses in the Galaxy. The results were also considered for contaminants, varying extinction, and varying metallicity. While the former alters the stellar parameters greatly, the effect of the latter is much smaller. 

From our results, we are able to discuss the star formation history of MB\,-C. Our major results are thus--

\begin{enumerate}
    \item MB\,-C is a star-forming region similar to OB associations in the Milky Way. The region can be separated into sub-regions, and is related to molecular clumps containing young, forming stars. YSOs and PMS stars form in molecular gas having densities between 17--200\,$M_{\odot}$\,pc$^{-2}$. Younger stars are on average found associated to molecular clouds with higher surface densities. 
    \item Molecular gas in MB\,-C shows filamentary structure, with the observed clump properties similar to those observed in the Galaxy. Molecular clumps, YSOs and young PMS candidates are closer in proximity to the filamentary structure than older PMS candidates, with older and young PMS populations appearing separated spatially. Star formation appears to proceed similar to star-forming regions observed in the Milky Way, with the low gas-to-dust, and CO content not hindering star formation. The bottleneck to forming stars may arise due to the conversion of atomic to molecular gas. 
    \item The discernible difference with respect to star formation in the Milky Way is the extinction towards YSOs, which is much lower in MB\,-C. The lower dust-to-gas ratio results in lower extinction towards the line of sight of protostars, which affects similarly the $X_{\rm CO}$ factor. However, the effect of lower dust (and corresponding shielding necessary to form H$_2$) does not appear to affect star formation, which is remarkably similar to the Milky Way.

\end{enumerate}

\section*{Acknowledgements}
V.M.K. acknowledges funding from CONICYT Programa de Astronomia Fondo Gemini-Conicyt as GEMINI-CONICYT 2018 Research Fellow 32RF180005. MR wishes to acknowledge support from  Universidad de Chile VID grant ENL22/18, from CONICYT(CHILE) through FONDECYT grant No1190684, and partial support from CONICYT project Basal AFB-170002. M.R. and H.S. wish to acknowledge support from CONICYT (CHILE) through FONDECYT grant 1140839, and partial support through project BASAL PFB-06. H.P.S. acknowledges financial support from the fellowship from CONICET and from SeCyT (C{\'o}rdoba, Argentina).
V.M.K. acknowledges useful comments made by Imogen Lawrie, Guillermo Blanc, Celia Verdugo, and the anonymous referee. This paper makes use of the following ALMA data:ADS/JAO.ALMA\#2015.1.1013.S. ALMA is a partnership of ESO (representing its member states), NSF (USA) and NINS (Japan), together with NRC (Canada), MOST and ASIAA (Taiwan), and KASI (Republic of Korea), in cooperation with the Republic of Chile. The Joint ALMA Observatory is operated by ESO, AUI/NRAO and NAOJ. This paper makes partial use of observations obtained at the Southern Astrophysical Research (SOAR) telescope, which is a joint project of the Minist\'{e}rio da Ci\^{e}ncia, Tecnologia, Inova\c{c}\~{o}es e Comunica\c{c}\~{o}es (MCTIC) do Brasil, the U.S. National Optical Astronomy Observatory (NOAO), the University of North Carolina at Chapel Hill (UNC), and Michigan State University (MSU). This work is based in part on observations made with the Spitzer Space Telescope, which is operated by the Jet Propulsion Laboratory, California Institute of Technology under a contract with NASA. This works uses in part data from Herschel. Herschel is an ESA space observatory with science instruments provided by European-led Principal Investigator consortia and with important participation from NASA. This work has utilised {\it Astropy} (http://www.astropy.org) a community-developed core Python package for Astronomy, the SEDFitter Python utility, and the STARLINK software TOPCAT.




\section*{Data availability}
The data underlying this article are available in the article and in its online supplementary material. The datasets were derived from sources in the public domain: ALMA science archive at http://almascience.nrao.edu/aq/; NOAO science archive at http://archive1.dm.noao.edu/; Spitzer archive at https://sha.ipac.caltech.edu/applications/Spitzer/SHA/; Herschel archive at https://www.cosmos.esa.int/web/hersch
el/data-products-overview. 




\appendix

\section{Comparison of YSO parameters}

\subsection{Comparison of SED fit parameters in this work to results from Chen et al. SED fits}

Chen et al. (2014) derived YSO masses based on fitting the models of Robitaille et al. (2006) to the infrared SEDs. The masses in those set of models are directly computed from the SED parameters using the Siess, Dufour \& Forestini (2000) models from stellar masses less than 7\,$M_{\odot}$, Bernasconi (1996) models are used for masses greater than 9\,$M{\odot}$, and masses in the 7--9\,$M_{\odot}$ range are interpreted between the two sets of models. The final masses in that paper were derived from solar metallicity isochrones.
The resulting fit parameters are compared to our masses in Table\,A1. 

\begin{figure}
\center 
\includegraphics[width=8cm]{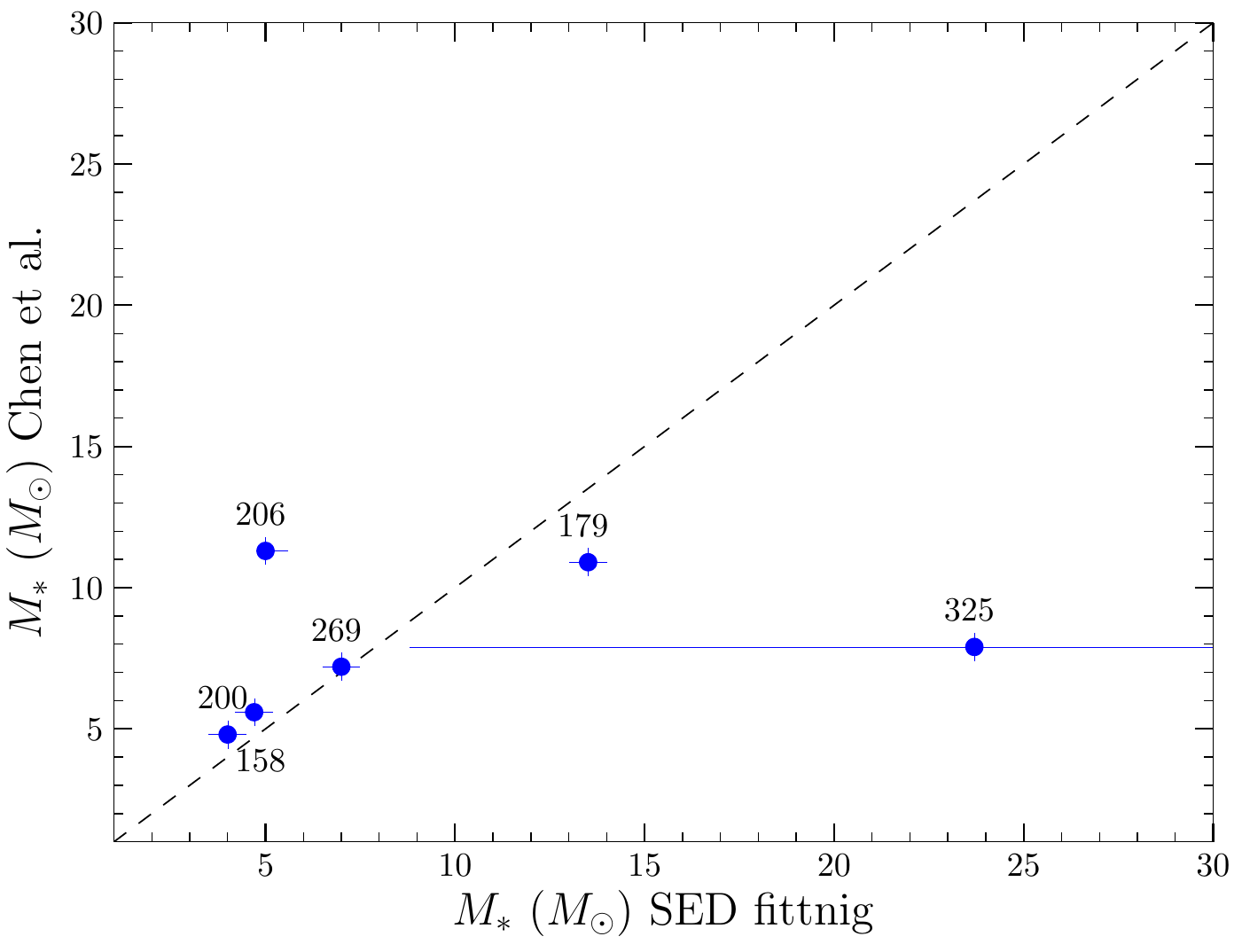}
\center
\caption{Comparison of stellar masses and the corresponding errors derived for YSO candidates identified in the mIR. The abscissa shows the masses estimated based on the best-fit model in this paper using Bressan et al. (2012) mass tracks, while the ordinate shows the same for the analysis performed by Chen et al. (2014). The range of masses from the SED fitting is shown as the error. Otherwise, a nominal error of 0.5\,$M_{\odot}$ is shown. The dashed line marks unity. Individual sources are labelled.}
\label{comparechenIR}
\end{figure} 

Based on the comparison (Fig.\,\ref{comparechenIR}), we find that our masses agree with the values quoted by Chen et al. (2014) for well-resolved sources in the infrared. Median differences for resolved sources is $\sim$0.2\,$M_{\odot}$. Unresolved source (\#179, \#325, \#206) present significantly larger differences in masses. Cause for these differences might be the higher angular resolution in the optical photometry employed in this work compared to Chen et al. (2014). A direct comparison in this scenario is not straightforward, and must take into account the choice of photometry used in the fits. Further causes may arise from the differences between the fitting methods (e.g. using a best-fit criteria compared to selecting the model with the most number of fits within a certain $\chi^2$ criterion; or weighted $\chi^2$ fit employed in Chen et al. 2014), and from the models themselves, which are extensively discussed in Robitaille (2017). A discussion of differences arising from the choice of models, or between optical and infrared imaging estimates is beyond the scope of the paper. Future efforts comparing past SED model fits to data obtained from telescopes with much higher angular resolution in the infrared are essential to shed light on this issue.

\begin{table*}
	\centering
\caption{A comparison of stellar parameters determined using mIR photometry fit to YSO models, and from optical photometry placed in the CMD at different extinction values.}

\begin{adjustbox}{max width=\textwidth}
\begin{threeparttable}
    \begin{tabular}{lcccccccc}
\hline
  \multicolumn{1}{c}{Name} &
  \multicolumn{1}{c}{Mass$_{\rm Opt.\,CMD}$} &
  \multicolumn{1}{c}{Age$_{\rm Opt.\,CMD}$} &
  \multicolumn{1}{c}{Mass$_{\rm IR\,SED}$} &
  \multicolumn{1}{c}{Age$_{\rm IR\,SED}$} &
  \multicolumn{1}{c}{$A_{V_{\rm IR\,SED}}$} &
  \multicolumn{1}{c}{Mass$_{\rm Chen\,et\,al.}$} &
  \multicolumn{1}{c}{Age$_{\rm Chen\,et\,al.}$} &
  \multicolumn{1}{c}{$A_{V_{\rm Chen\,et\,al.}}$}  \\
    \multicolumn{1}{c}{ } &
  \multicolumn{1}{c}{($M_{\odot}$)} &
  \multicolumn{1}{c}{(Myr)} &
  \multicolumn{1}{c}{($M_{\odot}$)} &
  \multicolumn{1}{c}{(Myr)} &
  \multicolumn{1}{c}{(mag)} &
  \multicolumn{1}{c}{($M_{\odot}$)} &
  \multicolumn{1}{c}{(Myr)} &
  \multicolumn{1}{c}{(mag)} \\
  \hline
\hline
  158 & 0.5 & $\lesssim$0.1 & 4.7 & 0.3 & 2.4 & 4.3 & 0.1 & 1.1 \\
  179 & 4.7 & 1.0 & 13.5 & $\lesssim$0.1 & 5.1 & 10.9 & 0.2 & 7.1 \\
  200 & 1.6 & 0.5 & 4.0 & 0.3 & 2.3 & 4.8 & 2 & 2.5 \\
  206$^a$ & 2.6 & 1.5 & 5.0 & 0.3 & 1.8 & 11.3 & 2 & 20.2 \\
  269 & 0.4 & $\lesssim$0.1 & 7.0 & $\lesssim$0.1 & 6.6 & 7.2 & 3 & 16.6 \\
  325 & 2.3 & 1.5 & 23.7 & $\lesssim$0.1 & 4.7 & 7.9 & $\lesssim$0.01 & 12.7 \\
\hline\end{tabular}
\begin{tablenotes}{\item[]$^{a}$ Contains a brighter visual companion within 1$''$ to the {\it Spitzer} source, and the cross-match may be incorrect.}
\end{tablenotes}
\end{threeparttable}
\end{adjustbox}
\end{table*}

\subsection{Comparison of SED fit parameters to  stellar properties estimated from optical photometry}

\begin{figure}
\center 
\includegraphics[width=8cm]{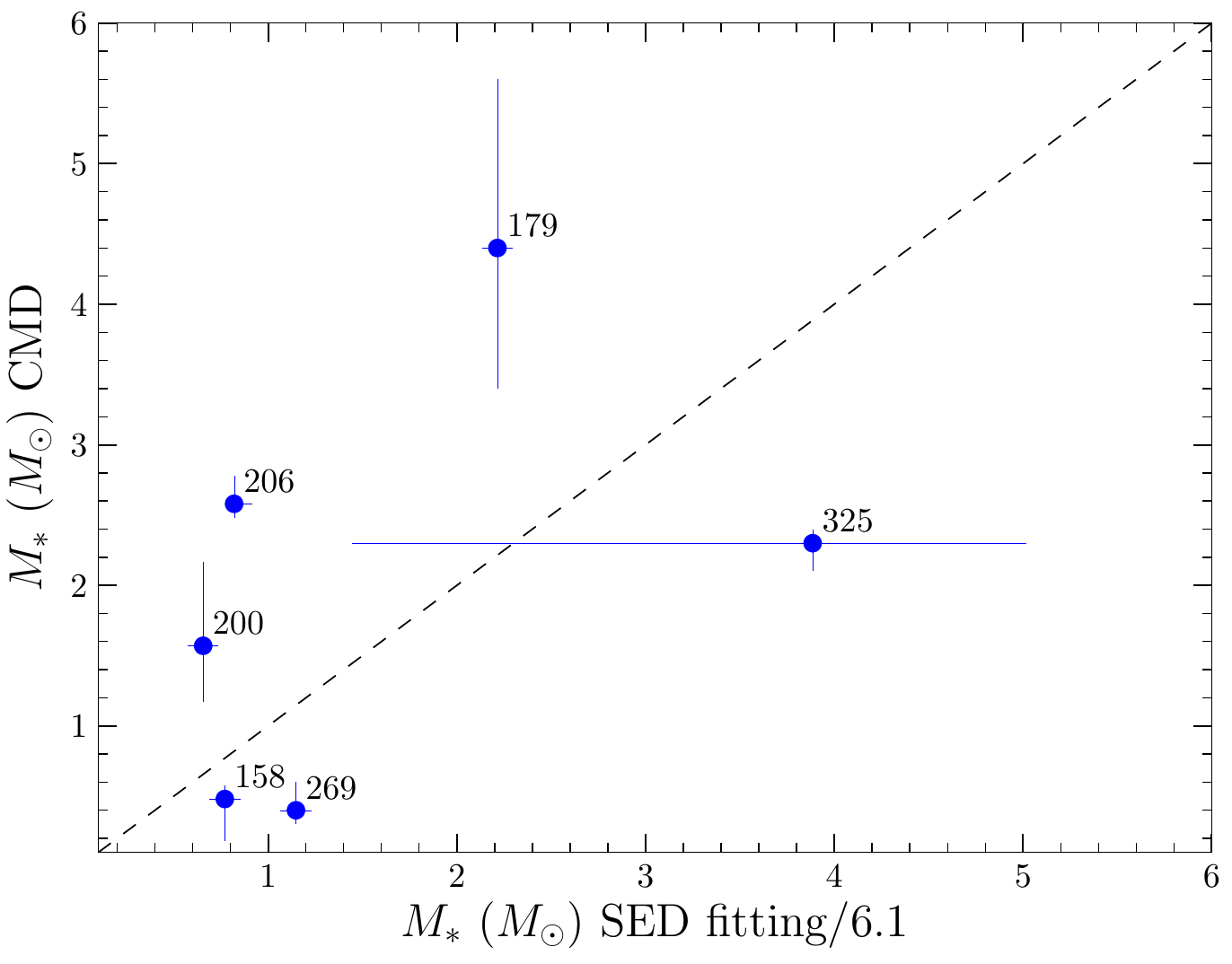}
\center
\caption{Comparison of stellar masses derived for YSOs identified in the mIR, and from optical photometry. The abscissa shows the masses estimated based on the best-fit YSO model in this paper using Bressan et al. (2012) stellar mass tracks over a factor of 6.1, while the ordinate shows the masses estimated using the same tracks based on the positions in the optical CMD. The range of masses from the SED fitting is shown as the error. Otherwise, a nominal error of 0.5\,$M_{\odot}$ is shown. The dashed line marks unity.}
\label{compare}
\end{figure}

The parameters derived from the SED fits to the YSO models might provide an independent corroboration of the values derived from the placing the optical photometry on the CMD. We compare the values derived from the latter, to the parameters determined based on the YSO models fit in this paper, and also Chen et al. (2014). The values from each analyses are listed in Table\,A1. It is apparent that the YSO masses from IR SED fits are significantly overestimated when compared to masses estimated by placing observed optical magnitudes in the CMD. The masses from the optical photometry are smaller than the IR SED fit masses by a median factor of 6.1 (Fig.\,\ref{compare}). This difference is hard to explain when considering simply the isochrones and metallicity, as the sets of isochrones and mass tracks used to estimate the results were the same in both scenarios. Similarly, the ages are significantly underestimated from the SED fits when compared to the optical photometric results. Although an estimate of the total mass from summing all detected optical sources over a composite mIR source might be made (for e.g., in Vaidya et al. 2009; Carlson et al. 2011), this value cannot be considered even if similar as a gauge for the accuracy of the mass derived from the SED parameter. This is because the temperature distribution may be affected in cases of blending multiple sources, making it hard to gauge the final effect on the estimated source mass (Robitaille 2017).

Another possible origin of this variation (when discounting differences due to isochrones themselves) might be the extinction estimates that were input into the two cases. Compared to the mean extinction derived from optical photometry ($A_V$=0.5\,mag), the extinction estimated by the YSO models is higher in most cases by at least values of  $\gtrsim$1\,mag. If the extinction towards individual sources in the optical is significantly underestimated, the objects would move towards the upper left in the CMD, following the reddening vector. This would match the prediction of underestimated masses, and ages when comparing the optical results to the SED fit results. Note that the estimated mass, and age of protostars is not linearly proportional to extinction. In Section\,4.5 we considered for this effect, and found for most values over $A_V$ of 2.5\,mag, the sources fell beyond the evolutionary tracks in the CMD. A glance at the extinction values derived by fitting against the YSO models reveals some nonphysical estimates of extinction. For example, stars with $A_V$ greater than 10\,mag are unlikely to be visible in the optical (assuming our limiting optical magnitude of $\sim$22.5\,mag) at a distance of 59\,kpc, unless they have absolute magnitudes greater than $-$6\,mag. This simplistic argument assumes a homogeneous distribution of dust, and that the optical and infrared object are the same and are resolved. It is noted that the range of extinction derived based on the YSO models spans 1.8-6.6\,mag, but the well-fit resolved sources have a median extinction $\sim$2.4\,mag, suggesting that the physical extinction might not be represented by the extreme values, but by values lower than 2.5\,mag.

The overestimation of extinction in the YSO SED fits may be due to not accounting for the correct dust-to-gas ratio (the reddening law was accounted for), which may lead to lower dust shielding implying there are smaller amounts of dust enshrouding the protostars. In a similar comparison, extinction values from the ratio of optical \ion{H}{I} gas lines, and $K$-band spectra (comparing infrared H$_2$ line ratios) were found to be an order of magnitude lower by Ward et al. (2017). The authors of that paper also found that the extinction scales proportionally with metallicity, and regions of low metallicity such as the Magellanic Bridge have protostars with lower amounts of dust surrounding them. This difference was attributed by the authors to the assumption that the optical observations trace the shallower regions of the protostars, while the infrared observations probe deeper into the circumstellar environment of the protostar. 
However, we note that H$_2$ infrared lines arise from the surfaces of molecular clumps that have been either shock or fluorescence excited (in protostars), but the ability to relate molecular gas, and dust is dependent on the mixing, local temperature, and other factors whose relative contribution is unclear (Ciurlo et al. 2019), and may not always be related to locally enhanced extinction due to the lack of excited molecular gas (e.g. Nadeau, Riopel \& Geballe 1991). Therefore, the assumption that the different wavelengths trace different optical depths may not be completely accurate in all scenarios. The overestimation of extinction values in the SED fits therefore lead to a systematic overestimation of stellar mass, and ages. We consider this effect, by re-fitting the SED models assuming a fixed extinction range between 0.1 and 1\,mag. This leads to an overestimate of the stellar mass by around 1-3 times the optically estimated stellar mass, and ages. Therefore, even accounting for extinction, some differences remain between the optically estimated stellar parameters to the SED fits. Future optical to IR flux-calibrated spectroscopy is vital to resolve this issue, allowing to estimate the extinction from multiple line ratios across the wavelength range, and from the slope of the spectra itself.

In addition to the factor of extinction, an important contributor to the differences between results from the two different wavelengths is the resolution of the imaging, which is nearly double that of the optical in the infrared. If a source is unresolved in the infrared, but resolved in the optical, then its total mass determined from fitting the SED to YSO models will be overestimated, as the infrared intensity will contain additional flux (and depending on the contaminant source, the SED at longer or shorter wavelengths might be altered). Since the YSO models themselves do not estimate the stellar mass or age, this effect on the final stellar parameters is hard to gauge. This suggests that multiplicity may be a viable cause for the differences observed, and future high angular resolution mid-infrared observations are necessary to resolve this issue.

We thus conclude our comparison by commenting on the fact that the diversity of values suggests a need for better resolution infrared and optical imaging to study protostars outside our Galaxy, in which the protostars at multiple wavelengths are similarly resolved (allowing to separate the stellar and envelope/disc components accurately) before judging the results of fitting models to SEDs. In addition, a more comprehensive study of extinction towards such sources, confirming that they are indeed lower their Galactic counterparts is only possible with optical to infrared flux calibrated spectroscopy, allowing extinction to be determined from multiple gas line ratios across the optical and infrared range, as well as the shape of the SED. Barring such data, we cannot determine precise estimates of YSO parameters, and it is essential not to over interpret results.  


\bsp	
\label{lastpage}
\end{document}